\documentclass[preprint2]{aastex63}

\usepackage{xfrac}
\usepackage{amsmath}
\usepackage{upgreek}

\newcommand\Rm{{\rm Rm} }

\newcommand\Pm{{\rm Pm} }
\newcommand\kf{k_{\rm f} }
\newcommand\SNr{\dot\sigma_{\rm sn}}

\newcommand\ESK{E_{\rm kin}}
\newcommand\EST{E_{\rm th}}

\newcommand{\vect}[1]{{{\mbox{\boldmath $#1$}}}}
\newcommand{\mathbfss}[1]{\textbf{\textsf{#1}}}
\newcommand\kpc{~ {\rm kpc}}
\newcommand\pc{~ {\rm pc}}
\newcommand\dx{ {\delta x}}
\newcommand\Myr{~ {\rm Myr}}
\newcommand\erg{~ {\rm erg}}
\newcommand\kms{~ {\rm km~ s}^{-1}}
\newcommand\BKM{{\sf BKMM4}}

\definecolor{midblue}{rgb}{0.0,0.4,0.7}
\definecolor{midgreen}{rgb}{0.1,0.6,0.3}
\definecolor{mypurple}{rgb}{0.7,0.3,0.8}

\received{October 4, 2020}
\revised{\today}
\accepted{}
\submitjournal{ApJL}

\shorttitle{Small-scale dynamo in the ISM}
\shortauthors{Gent et al.}

\begin{document}

\title{Small-Scale Dynamo in Supernova-Driven Interstellar Turbulence}

\correspondingauthor{Maarit K\"apyl\"a}
\email{Email: frederick.gent@aalto.fi, mordecai@amnh.org,\\ maarit.kapyla@aalto.fi, nishant@iucaa.in}

\author[0000-0002-1331-2260]{Frederick A. Gent}
\affiliation{
Astroinformatics, Department of Computer Science, Aalto University, PO Box 15400, FI-00076 Espoo, Finland
 }
\affiliation{
    School of Mathematics, Statistics and Physics,
      Newcastle University, NE1 7RU, UK 
 }

\author[0000-0003-0064-4060]{Mordecai-Mark {Mac Low}}
\affiliation{
  Department of Astrophysics, {79th Street at Central Park West, }American Museum of Natural History,
  New York, NY 10024, USA
}
\affiliation{
{Center for Computational Astrophysics, {162 Fifth Avenue, }Flatiron Institute, New York,
NY 10010, USA} 
}

\author[0000-0002-9614-2200]{Maarit J. K\"apyl\"a}
\affiliation{
Astroinformatics, Department of Computer Science, Aalto University, PO Box 15400, FI-00076 Espoo, Finland
}
\affiliation{
Max Planck Institute for Solar System Research, Justus-von-Liebig-Weg 3, 37707 G\"ottingen, Germany
}
\affiliation{
    Nordic Institute for Theoretical Physics,
      Roslagstullsbacken 23, 106 91 Stockholm, Sweden 
}

\author[0000-0001-6097-688X]{Nishant K. Singh}
\affiliation{
Inter-University Centre for Astronomy \& Astrophysics, Post Bag 4, Ganeshkhind, Pune 411 007, India
}
\affiliation{
Max Planck Institute for Solar System Research, Justus-von-Liebig-Weg 3, 37707 G\"ottingen, Germany
}


\begin{abstract}
Magnetic fields grow quickly even at early cosmological times, suggesting the
action of a small-scale dynamo (SSD) in the interstellar medium of galaxies.
Many studies have focused on idealized turbulent driving of the SSD. 
Here we simulate more realistic supernova-driven turbulence to determine
whether it can drive an SSD.
Magnetic field growth occurring in our models appears inconsistent with simple
tangling of magnetic fields, but consistent with SSD action, reproducing and
confirming models by \citet{BKMM04} that did not include physical resistivity
$\eta$.
We vary $\eta$, as well as the numerical resolution and supernova rate,
$\dot\sigma$, to delineate the regime in which an SSD occurs.
For a given $\dot\sigma$ we find convergence for SSD growth rate with
resolution of a parsec.
For $\dot\sigma\simeq\dot\sigma_{\rm sn}$, with $\dot\sigma_{\rm sn}$ the solar
neighbourhood rate, the critical resistivity below which an SSD occurs is
$0.005>\eta_{\rm crit}>0.001\,\rm kpc^{-1}\,\rm km~ s^{-1}$, and this
increases with the supernova rate.
Across the modelled range of 0.5--4~pc resolution we find that for
$\eta<\eta_{\rm crit}$, the SSD saturates at about 5\% of kinetic energy
equipartition, independent of growth rate.
In the range $0.2\,\dot\sigma_{\rm sn}\leq\dot\sigma\leq8\,\dot\sigma_{\rm sn}$
growth rate increases with $\dot\sigma$.
SSDs in the supernova-driven interstellar medium commonly exhibit erratic
growth.
\end{abstract}
\keywords{dynamo --- magnetohydrodynamics (MHD) --- ISM: supernova remnants --- ISM: magnetic fields --- turbulence}

\section{Introduction}\label{sec:intro}

     {We here study the}
 small-scale dynamo (SSD) in the interstellar medium (ISM).
 SSD acts at small eddy scales of turbulence, driving magnetic field
 growth at correspondingly short {timescales}.
 {The large-scale dynamo (LSD) with much longer turnover times generates
 magnetic fields ordered on {kiloparsec} scales.}
 Hence, capturing LSD alongside the faster growing modes
 of SSD in simulations is computationally challenging.
 However, {interaction between SSD and LSD modes likely fundamentally}
 {determines the} evolution and structure of the magnetic field.

 Many simulations of supernova- (SN)-driven turbulence with realistic vertical
 stratification \citep[e.g.,][]{deAvillez:2005,PO07,Hill:2012a,HI14} have no
 mechanism to induce LSD, such as rotation and shear.
 Strong ordered magnetic field effects are modelled by
 imposition of a background, typically uniform, magnetic field.
 {Some large-scale models do seek to include LSD 
 \citep[e.g.,][]{Korpi:1999b,Gressel:2008,HWK09,WA09,Pakmor17,
 GE20}, but show no SSD, or appear to find SSD 
 within the context of halo-disk
 scale flows \citep[e.g.,][]{RT16,SBADMN19},  but capture no LSD.}
\citet[][with additional analysis by \citealt{EGSFB16}]{Gent:2013b}
 appear to include an {SSD with} {an} {LSD}.
 To confirm this and determine its effect on LSD, we must understand the
 properties of the SSD.
     
 Any magnetic noise produced by tangling {of a large-scale field}
 will also grow exponentially {if an LSD is} present.
 This noise {can play} an important role in quenching the LSD.
 We need to discriminate this effect from an SSD.   

 Previous experiments ({\citealp[e.g.,][hereafter \BKM]{BKMM04};
 \citealp{BalKim05,MacLow:2005}})
 examined the SN-driven SSD.
 {The limited  resolution} {study} of \BKM\ did not allow
 demonstration of {solution} convergence.
     Furthermore, they imposed a uniform background field and
      implemented no physical resistivity or viscosity.
 We shall show that the amplification of their field is a result
 of SSD action and not just tangling of the field.

 In this {L}etter we {first} compare the SSD to tangling in an idealized
 simulation (Sect.~\ref{sec:ssd-tang}){.
 We then describe our models of SN-driven turbulence for demonstrating the
 action of SSD (Sect.~\ref{sec:model}).} 
 {Simulations} use the {\sc Pencil Code}\footnote{
 \href{https://github.com/pencil-code}{https://github.com/pencil-code}}.
 A broad resolution and parameter study allows us to {show numerical}
 {convergence and} {determine} the critical resistivity for excitation
 of an SSD, {which we follow} to saturation (Sect.~\ref{sec:results}).
 This provides objective criteria {for the action} of SSD in simulations
 \citep[such as][]{Gent:2013b,GE20,SBADMN19}.
 Finally, we conclude in Sect.~\ref{sec:conc}.
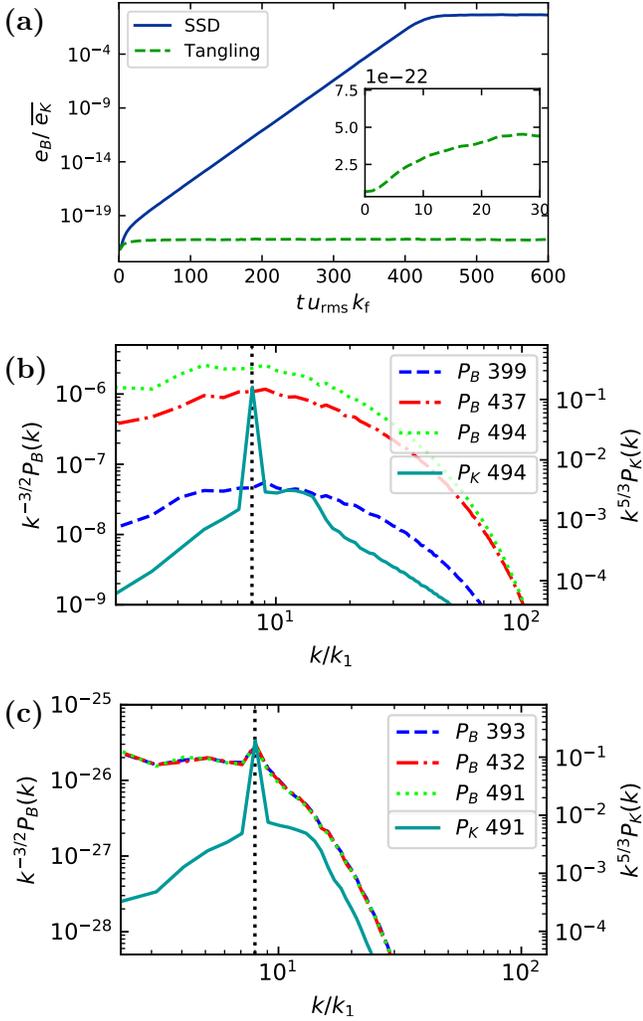
\begin{figure}
  \includegraphics[trim=0.00cm 0.3cm 0.0cm 0.0cm, clip=true,width=0.91\columnwidth]{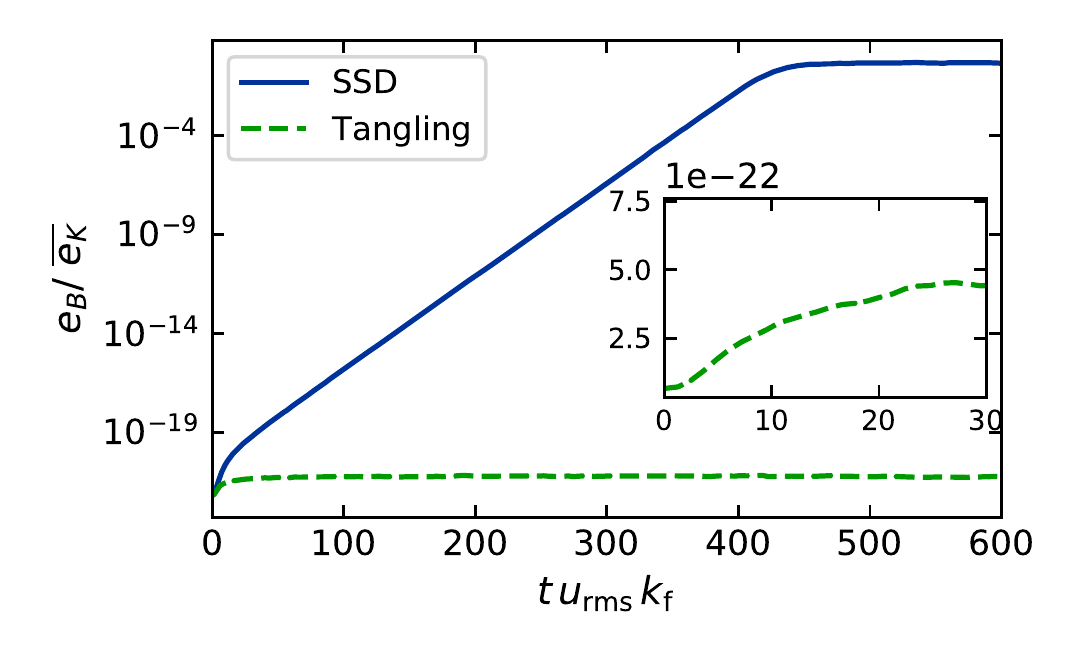}
  \includegraphics[trim=0.25cm 0.3cm 0.5cm 0.1cm, clip=true,width=1.0\columnwidth]{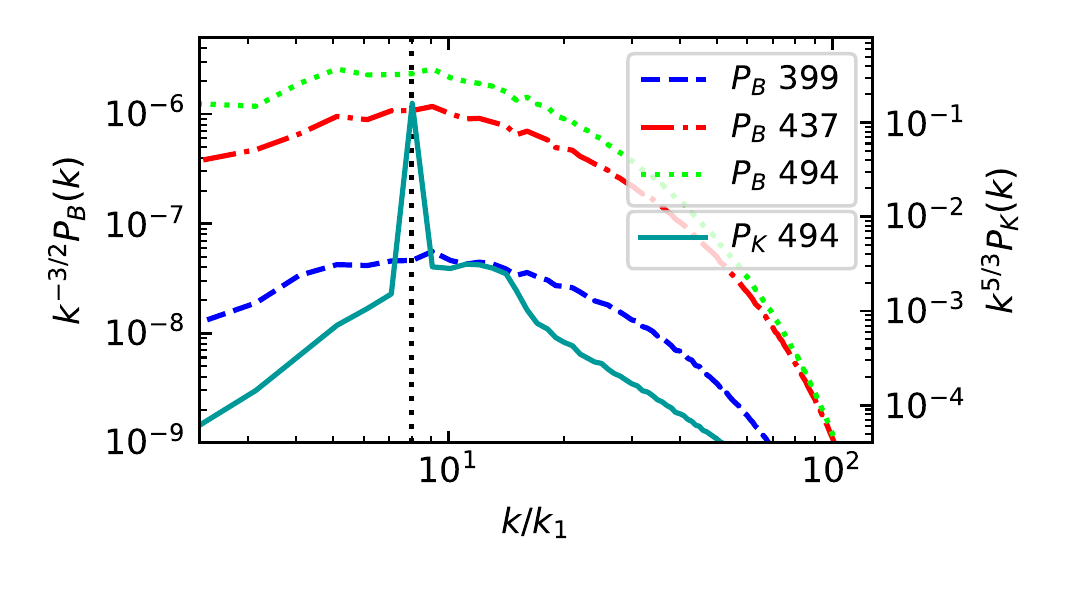}
  \includegraphics[trim=0.35cm 0.6cm 0.5cm 0.3cm, clip=true,width=1.0\columnwidth]{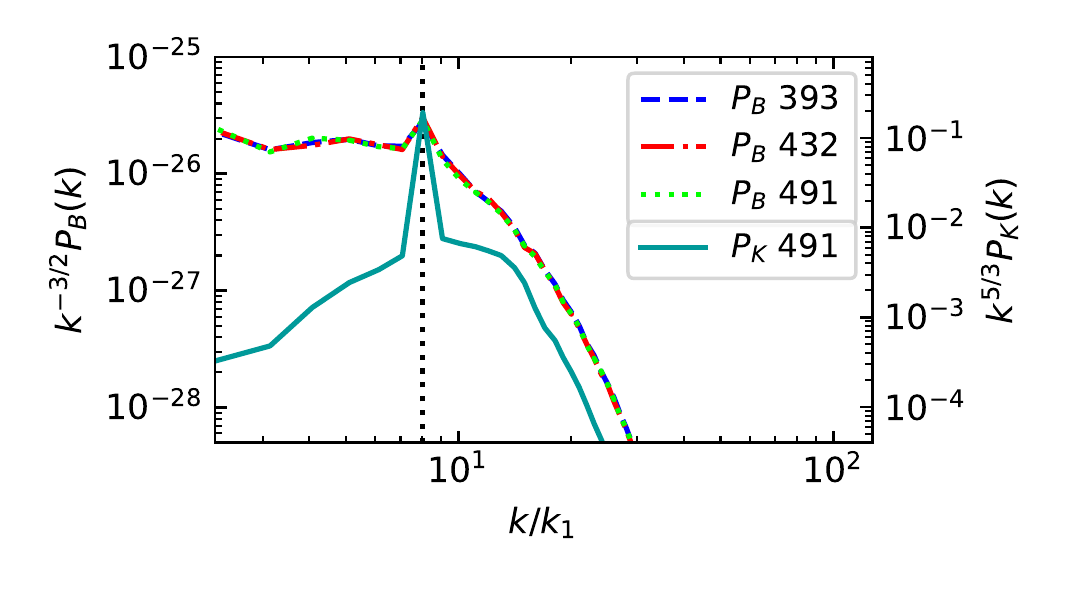}
  \begin{picture}(0,0)(0,0)
    \put(0,387){{\sf\bf{(a)}}}
    \put(0,255){{\sf\bf{(b)}}}
    \put(0,125){{\sf\bf{(c)}}}
  \end{picture}
\caption{
 (a) Mean magnetic energy density, $e_B$, with nonhelical random forcing,
 scaled to time-averaged kinetic energy density, $\overline{e_K}$.
 Inset: early zoom-in of linear growth of tangled field.
 Time is normalised by eddy turnover time, $1/\kf {u_{\rm rms}}$.
 Compensated power spectra {of kinetic energy $P_k$ and magnetic
 energy $P_B$} for (b) SSD and (c) tangling, at times given in the
 legends.  Kinetic energy uses the right-hand axes.
 Forcing scale, $\kf/k_1=8$: vertical dotted line.
\label{fig:tangling}}
\end{figure}

\section{Disentangling the dynamo} \label{sec:ssd-tang}

 {Previous SSD studies have examined Pm dependence with 
 {stochastic} 
 nonhelical forcing, including {at} high Mach number \citep[e.g.,][]{ 
 HBD03,HBD04,Haugen:2004M,FCSBKS11,FSBS14}.} 
 {Here we specifically seek t}o illustrate
 differences between tangling and SSD.
 Nonhelical random forcing is applied at wavenumber $\kf/k_1=8$ to
 $256^3$ zone, $2\pi$-periodic, isothermal boxes
     {with viscosity $\nu=5\cdot10^{-3}$}.
 The lowest wavenumber in the domain is $k_1=1$ and the largest is the Nyquist
 frequency $k/k_1 = 128$.
 The imposed uniform field has $e_B\simeq6\cdot10^{-22}~\overline{e_K}$, where
 $\overline{e_K}$ is the time-averaged kinetic energy density.
 
 Two simulations are distinguished by use of dimensionless
 resistivity $\eta=10^{-4}$
 and $\eta=2\cdot10^{-3}$.
  Respectively, these yield
     {magnetic Reynolds number} $\Rm=150$, with {magnetic Prandtl} number $\Pm = 50$,
 exciting {an} SSD and $\Rm=7.4$,
 with $\Pm=2.5$, inhibiting the dynamo so that amplification is limited to
 tangling of the imposed field.

 Figure\,\ref{fig:tangling}\,(a) shows the SSD growing exponentially in just
 over 400 eddy turnover times; see \cite{ZRS83} for SSD properties and
 excitation conditions.
 Tangling induces only linear growth (see inset), saturating just above
 the imposed field energy within 50 turnover times.

 {Figures~\ref{fig:tangling}(b) and~\ref{fig:tangling}\,(c) show
 compensated power spectra for both cases.}
 Magnetic energy spectra are compensated {for by} Kazantsev's $k^{3/2}$
 {power law} \citep{Sch02,BS14}, and kinetic energy {by} Kolmogorov's
 $k^{-5/3}$.
 The forcing scale $\kf$ is {prominent} in
 the magnetic energy {spectra of the tangling} but
 {not} {in the magnetic spectra of the SSD}.
 {For SSD the range with Kazantsev power law (horizontal) extends to scales
 smaller than the forcing scale (Figure~\ref{fig:tangling}\,b), and during the
 kinematic phase the magnetic energy peak is at $k/k_1\simeq9$.}
 {For tangling (Figure~\ref{fig:tangling}\,c) the Kazantsev
 {scaling} applies only at $k/k_1<\kf$.} 
 Thus, in the SSD, kinetic energy {in the Kolmogorov cascade} transfers to
 {magnetic energy} 
 at these scales, inducing an inverse Kazantsev {cascade}
 at scales {smaller than} $\kf$, while tangling transfers energy only at
 scales between $\kf$ and the scale of the imposed field.
 
\section{{Supernova-driven} turbulence model design} \label{sec:model}

 {Our} SN-driven turbulence models exclude large-scale magnetic field
 dynamics by {omitting global-scale} rotation, shear, and stratification.
 Our simulation domain is a periodic cube of length 256 pc and zone size
 $\dx=0.5$, 1, 2 or 4$\pc${, except for {our} direct comparisons with
 \BKM, { which have} domains {of} 200 pc and $\dx=0.78$, 1.56 and
 3.12$\pc$ (units henceforth assumed)}.
 {Our fiducial models exclude tangling of an imposed field as a source
 of magnetic amplification, by applying a random 10~nG initial field.
 Transient dissipation
     prior to hydrodynamic steady state and dynamo onset yields
a turbulent seed field of about 1~nG.
 For models reproducing \BKM\ this seed is substituted by
 a uniform $10$~nG background field as applied by \BKM.}

\begin{figure*}
  \centering
  \includegraphics[trim=0.4cm 0.5cm 0.3cm 0.3cm, clip=true,width=0.87\linewidth]{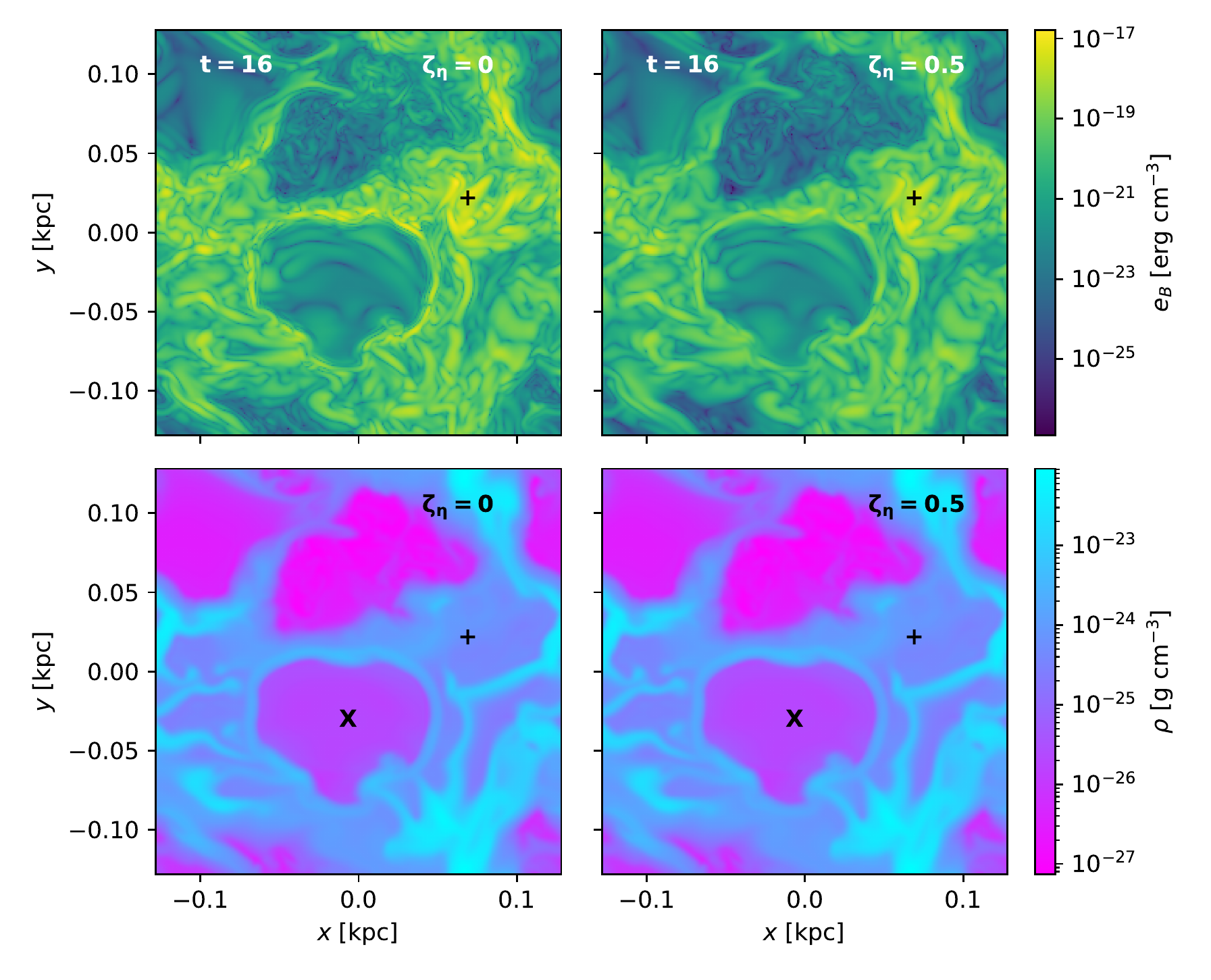}
  \begin{picture}(0,0)(0,0)
    \put(-395,205){{\sf\bf{(a)}}}
    \put(-225,205){{\sf\bf{(b)}}}
    \put(-395, 35){{\sf\bf{(c)}}}
    \put(-225, 35){{\sf\bf{(d)}}}
  \end{picture}
\caption{{2D slices
of (a, b) magnetic energy density $e_B$ and (c, d) gas density
with shock dependent resistivity, $\zeta_\eta$, as indicated.
The site of the most recent SN is indicated by X (c, d).
$\dx=1$\,pc, $\eta=10^{-4}$ and $\nu=10^{-3}\kms\kpc^{-1}$. }
\label{fig:eb-slice}}
\end{figure*}
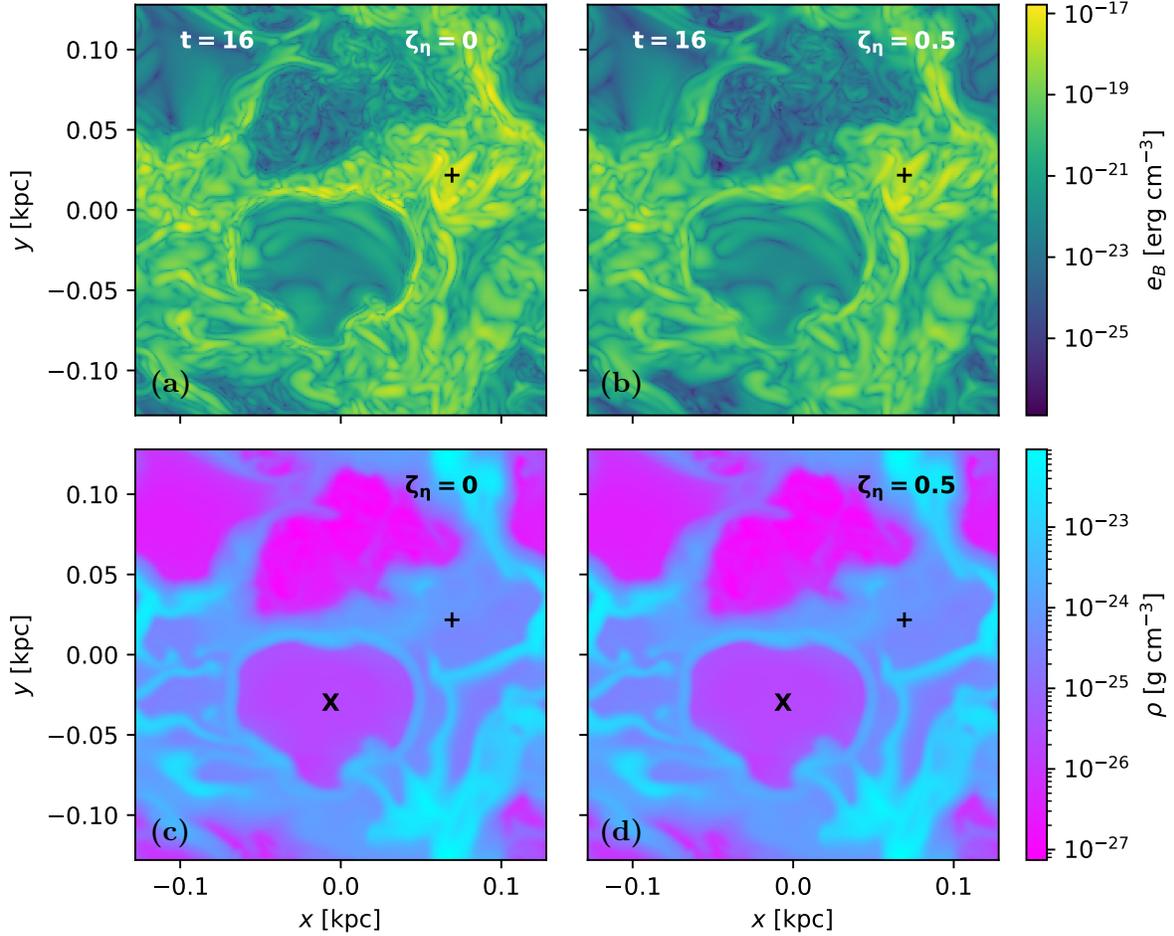

 We solve the system of non-ideal, compressible, non-isothermal MHD equations
  \begin{eqnarray}
  \label{eq:mass}
    \frac{D\rho}{Dt} &=& 
    -\rho \vect\nabla \cdot \vect{u}
    +\vect\nabla \cdot\zeta_D\vect\nabla\rho,
  \end{eqnarray}
  \begin{eqnarray}
  \label{eq:mom}
    \rho\frac{D\vect{u}}{Dt} &=& 
    \vect\nabla{\ESK\sigma}
    -\rho c_{\rm s}^2\vect\nabla\left({s}/{c_{\rm p}}+\ln\rho\right)
    +\vect{j}\times\vect{B}
    \nonumber\\
    &+&\vect\nabla\cdot \left(2\rho\nu{\mathbfss W}\right)
    +\rho\,\vect\nabla\left(\zeta_{\nu}\vect\nabla \cdot \vect{u} \right)
    \nonumber\\
    &+&\vect\nabla\cdot \left(2\rho\nu_3{\mathbfss W}^{(3)}\right)
  {-\vect u\vect{\nabla}\cdot\left(\zeta_D\vect{\nabla}\rho\right)},
  \end{eqnarray}
  \begin{eqnarray}
  \label{eq:ent}
    \rho T\frac{D s}{Dt} &=&
     \EST\dot\sigma +\rho\Gamma
    -\rho^2\Lambda +\eta\mu_0\vect{j}^2 
    \nonumber\\
    &+&2 \rho \nu\left|{\mathbfss W}\right|^{2}
    +\rho\,\zeta_{\nu}\left(\vect\nabla \cdot \vect{u} \right)^2
    \nonumber\\
    &+&\vect\nabla\cdot\left(\zeta_\chi\rho T\vect\nabla s\right)
    +\rho T\chi_3\vect\nabla^6 s
    \nonumber\\
    &-& {c_{\rm{v}}\,T \left(
    \zeta_D\nabla^2\rho + \vect\nabla\zeta_D\cdot\vect\nabla\rho\right)},
  \end{eqnarray}
  \begin{eqnarray}
  \label{eq:ind}
    \frac{\partial \vect{A}}{\partial t} &=&
    \vect{u}\times\vect{B}
    +\eta\vect\nabla^2\vect{A}
    +\eta_3\vect\nabla^6\vect{A},
  \end{eqnarray}
 with the ideal gas equation of state closing the system.
 Most variables take their usual meanings.
 Terms containing $\zeta_D{=2},\,\zeta_\nu=5$ and $\zeta{_\chi=2}$
 {are applied to all ISM models and} resolve shock discontinuities with
 artificial diffusion of mass, momentum, and energy proportional to shock
 strength \citep[see][for details]{GMKSH20}.
 {Equations~\eqref{eq:mom} and \eqref{eq:ent} include terms with $\zeta_D$}
 {to} {provide momentum and energy conserving corrections for} {the}
 {artificial mass diffusion applying in Equation~\eqref{eq:mass}.}
 {In previous work \citet{Gent:2013b} we have used a formalism that
 included artificial diffusion in vector potential at shocks.
 In Figure~\ref{fig:eb-slice} we show comparative slices of the magnetic
 energy and gas density with and without resistive shock diffusion
 $\zeta_\eta$.
 With $\zeta_\eta>0$ (Fig.~\ref{fig:eb-slice} b) magnetic energy is reduced in the remnant shell relative to Figure~\ref{fig:eb-slice} a, where compression actually enhances it.}
 Since the magnetic field is well resolved in either case, as also
 shown by the magnetic energy spectra below, and the simulation is
 numerically stable without it, this extra artificial
 diffusion is unnecessary.

 In both models a concentration of magnetic energy, marked with $+$ in
 Figure\,\ref{fig:eb-slice}, has below average gas density.
 {This snapshot reflects the overall behaviour of the system, in which
 magnetic field amplification also occurs independently of shock compression.
 As Figure\,\ref{fig:eb-slice} shows, SN shock fronts do compress and amplify
 the magnetic field, resulting in strong local and instantaneous correlation
 of the field and density.
 However, on global and long-term scale, this is not the dominant mechanism
 for the dynamo, which operates just as effectively in the non-shocked, more
 diffuse regions, as is also indicated by this figure.
 This is based on the amplification factor due to compression being estimated
 $\lesssim2$, taken as density fluctuations to power $4/3$, while the magnetic
 energy is amplified by 4--6 orders of magnitude.}

  {Unlike past} experiments \citep{Gent:2013b,Gent:2013a,GMKSH20},
 thermal diffusivity $\chi$ is {also} omitted, as the artificial diffusivities
 {chosen} are adequate to ensure numerical stability.
 {The} physical effects of thermal conductivity can be expected to be
 relevant only at the unresolved or marginally resolved Field length defined
 by \citet[][named after George Field, not the magnetic field]{BM90}.
 Terms containing $\nu_3,\,\chi_3$ and $\eta_3$ apply sixth-order hyperdiffusion
 to resolve grid-scale instabilities \citep[see, e.g.,][]{ABGS02,HB04}, {
 with mesh Reynolds number set to be $\simeq1$ for each $\dx$}.

 {The simplified isothermal model considered in
Sect.\,\ref{sec:ssd-tang} solves only Equations~{\eqref{eq:mass},}
 \eqref{eq:mom} and~\eqref{eq:ind}, without the shock-dependent diffusion or
 hyperdiffusion terms, and while setting
 $\vect{B}=\vect\nabla\times\vect{A}+\vect{B}_{\rm imposed}$.}

 {In the ISM simulations} SNe are exploded at {uniform} random positions
 at a Poisson rate $\dot\sigma$ {scaled by} the solar neighborhood
 value $\SNr\simeq 50\kpc^{-3}\Myr^{-1}$.
 Explosions inject $\EST = 10^{51}\erg$ thermal energy, except in
 dense regions, where a proportion {($<5\%$) may be} kinetic $\ESK$ 
 \citep[see][]{GMKSH20}.
 {Models with common $\dot\sigma$ have the same timing and location of
 explosions.}
 Non-adiabatic heating $\Gamma$ and cooling $\Lambda (T)$ are included
 \citep{Gent:2013b} following \citet{Wolfire:1995} and \citet{Sarazin:1987}.

\begin{figure}
  \centering
  \includegraphics[trim=0.1cm 0.2cm  0.3cm 0.2cm, clip=true,width=\columnwidth]{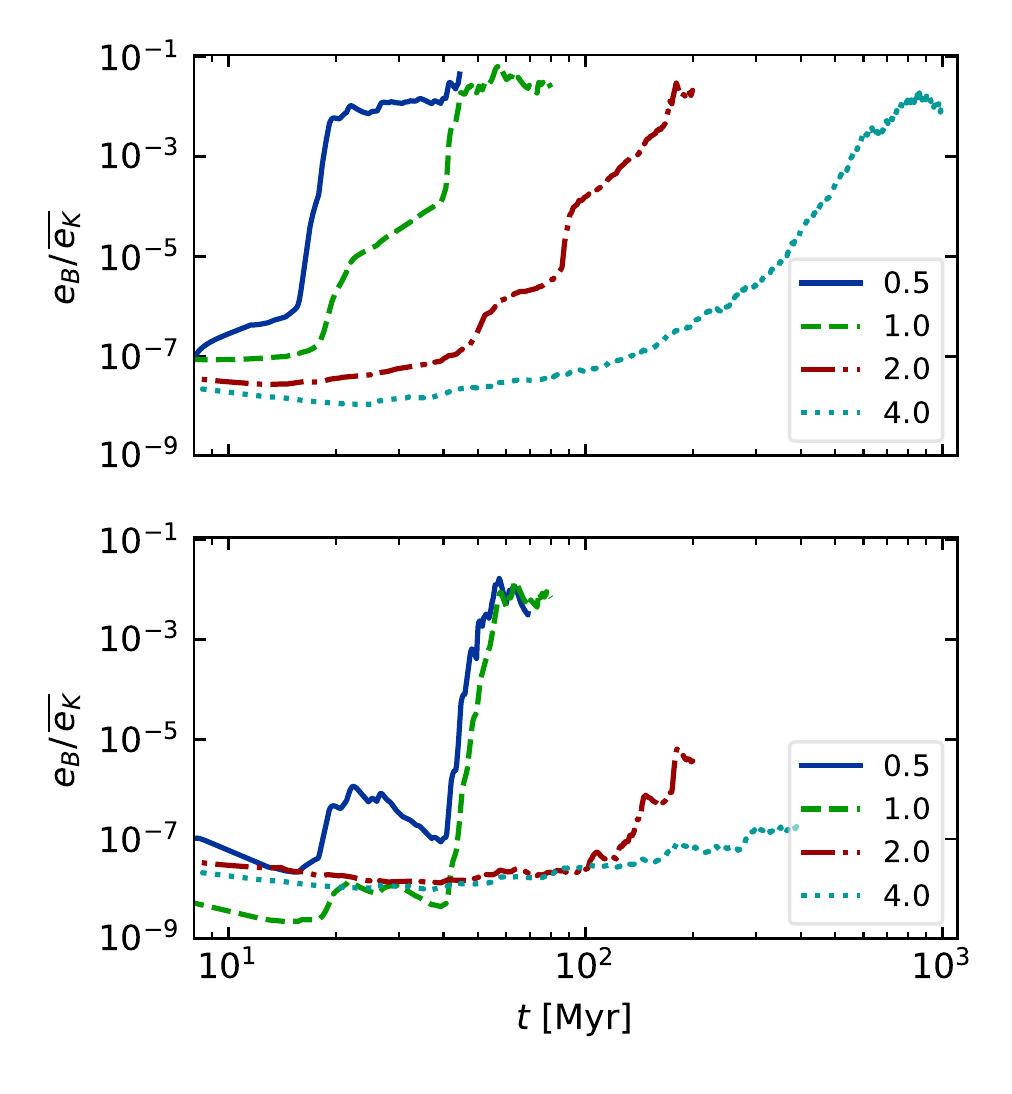}
  \includegraphics[trim=0.1cm 1.0cm  0.2cm 0.2cm, clip=true,width=\columnwidth]{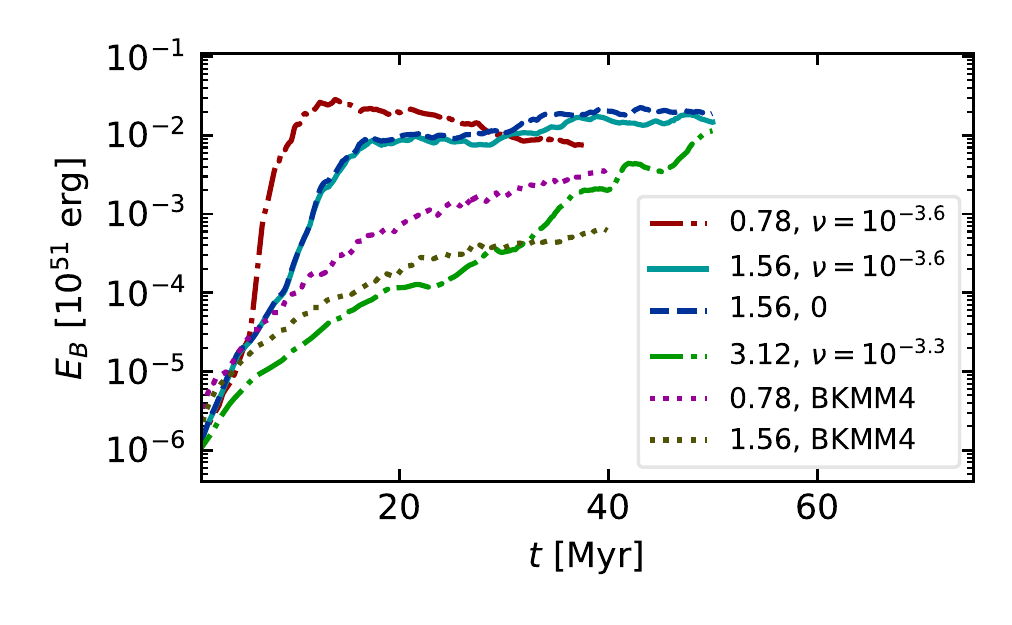}
  \includegraphics[trim=0.1cm 0.5cm  0.2cm 0.2cm, clip=true,width=\columnwidth]{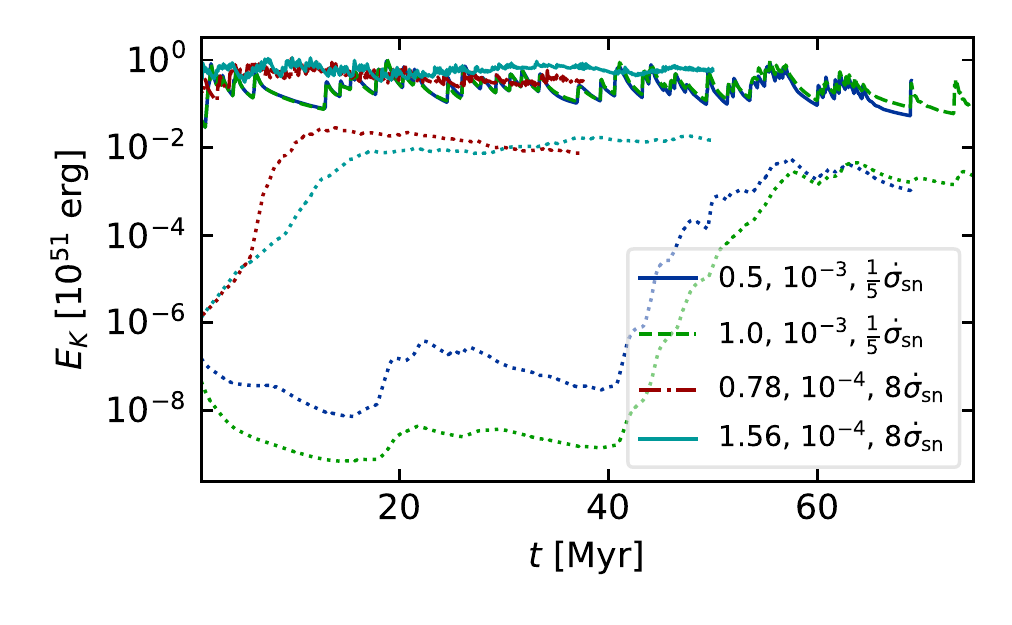}
  \begin{picture}(0,0)(0,0)
    \put(-120,520){{\sf\bf{(a)}}}
    \put(-120,395){{\sf\bf{(b)}}}
    \put(-120,255){{\sf\bf{(c)}}}
    \put(-120,133){{\sf\bf{(d)}}}
  \end{picture}
\caption{
 Magnetic energy density shown for resolutions $\dx$ given in the legends for
 models with resistivity (a) $\eta=10^{-4}$, and (b) $\eta=10^{-3}$, scaled by
 the time-averaged kinetic energy density $\overline{e_K}${. 
 Total magnetic energy, $E_B$} scaled by $\EST$, (c) matching \BKM\ {for}
 {$\dx$ and viscosity $\nu$} included in the legend and 
 {(d) kinetic energy, $E_K$, for $\dx$, $\eta$ and $\dot\sigma$ given in
 the legend with corresponding magnetic energy (dotted).}
\label{fig:eb-res}}
\end{figure}
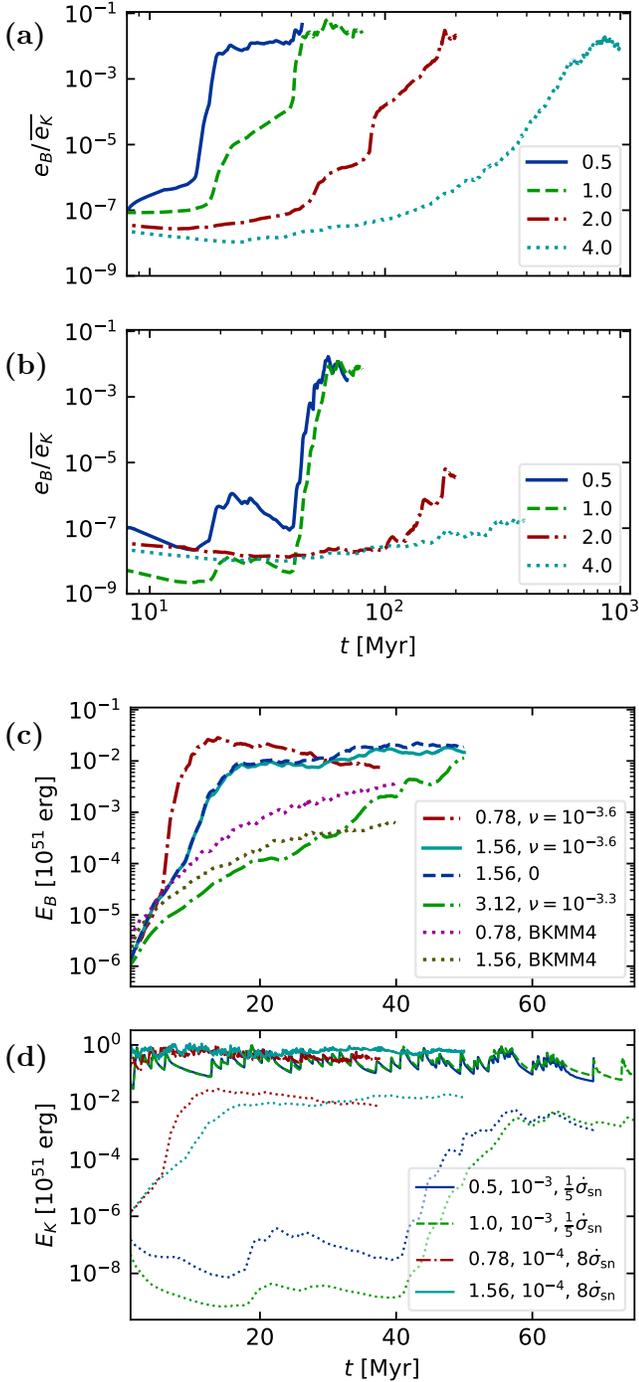

{
        {To understand} 
  the effects of purely numerical diffusivity, we also run an ideal MHD model
 with} $\eta=0$ {and $\nu=0$.}
 We determine how low a physical resistivity $\eta$ can be resolved by varying
 it from $10^{-5}$ to $10^{-3}\kpc\kms$ (units assumed henceforth).
 {We also test the effect of $\Pm=\nu/\eta$, varying $\nu$ with 
 $\eta=10^{-4}$ or varying $\eta$ with $\nu=10^{-3}$.}
 {Our direct comparison with the results of \BKM\ uses $\Pm=2.5$, apart
 from one run using $\eta=\nu=0$.}
 
\begin{figure*}
  \includegraphics[trim=0 1.2cm 0 0.2cm,clip=True,width=0.48\textwidth]{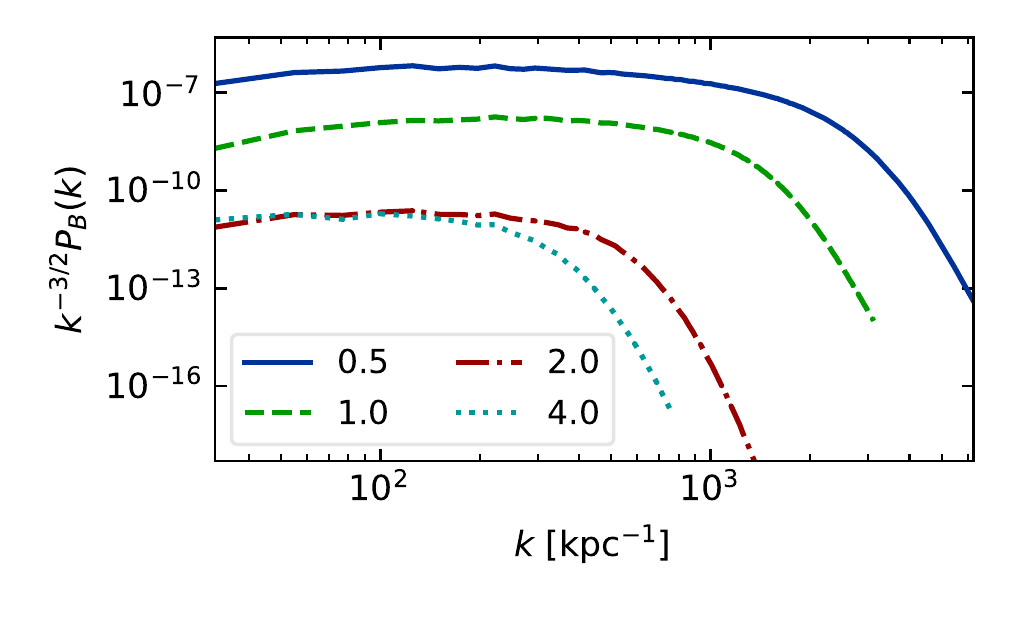}
  \includegraphics[trim=0 1.2cm 0 0.2cm,clip=True,width=0.48\textwidth]{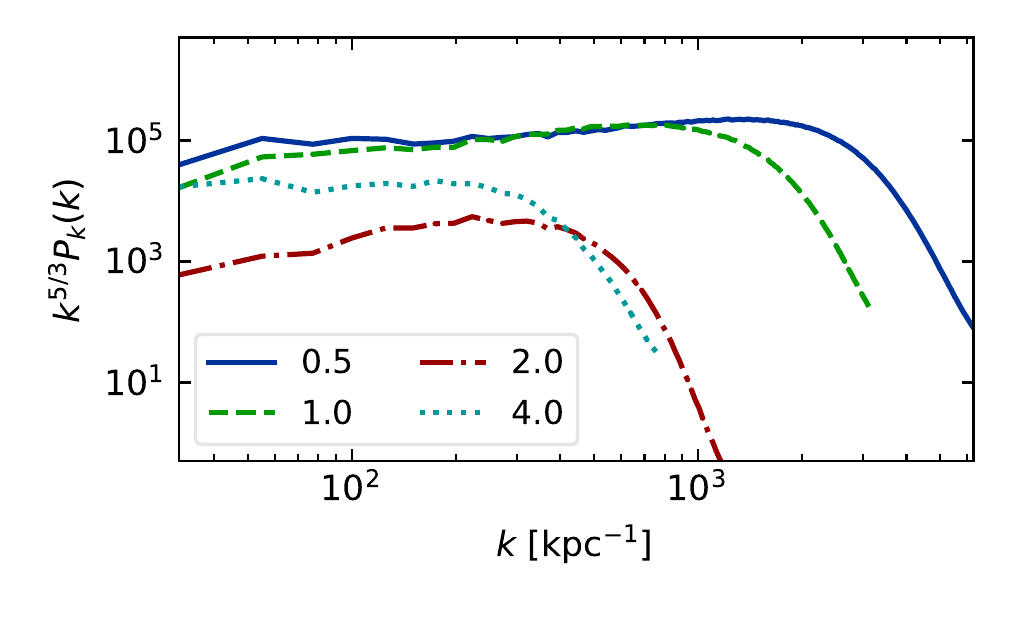}\\
  \includegraphics[trim=0 1.2cm 0 0.2cm,clip=True,width=0.48\textwidth]{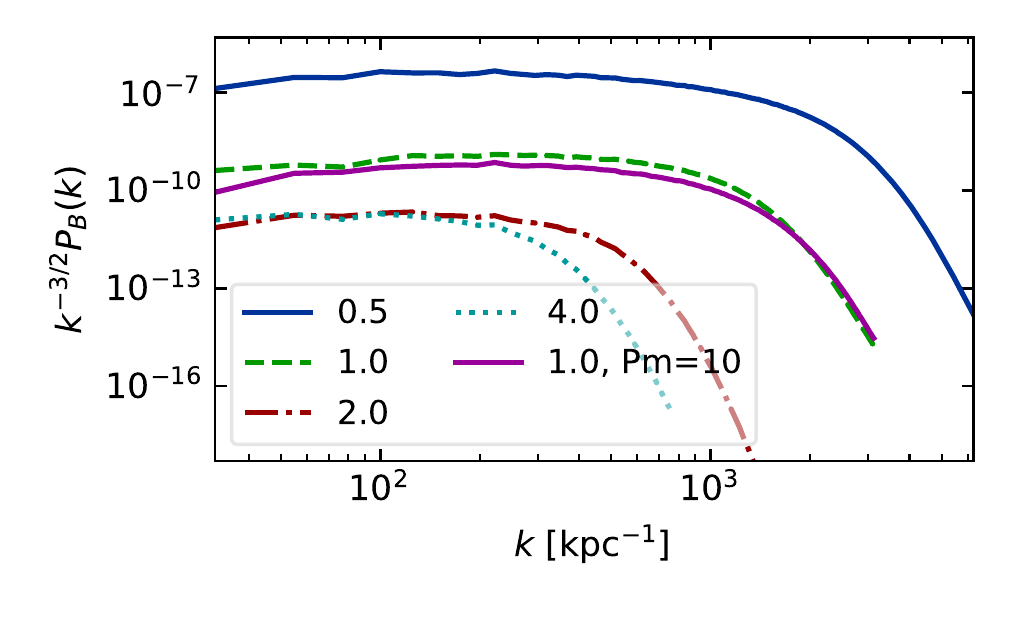}
  \includegraphics[trim=0 1.2cm 0 0.2cm,clip=True,width=0.48\textwidth]{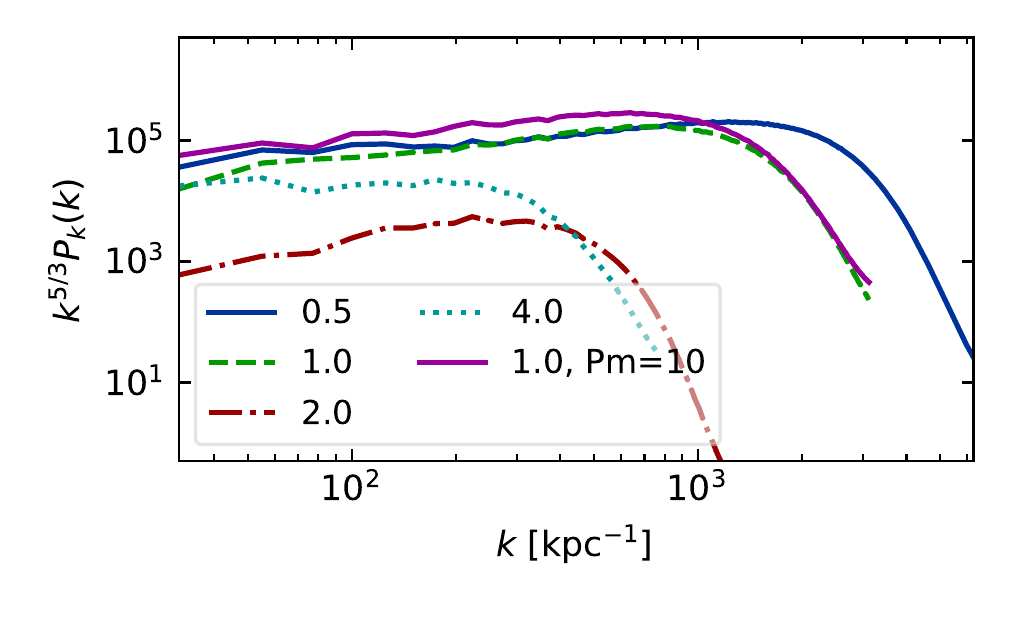}\\
  \includegraphics[trim=0 1.2cm 0 0.2cm,clip=True,width=0.48\textwidth]{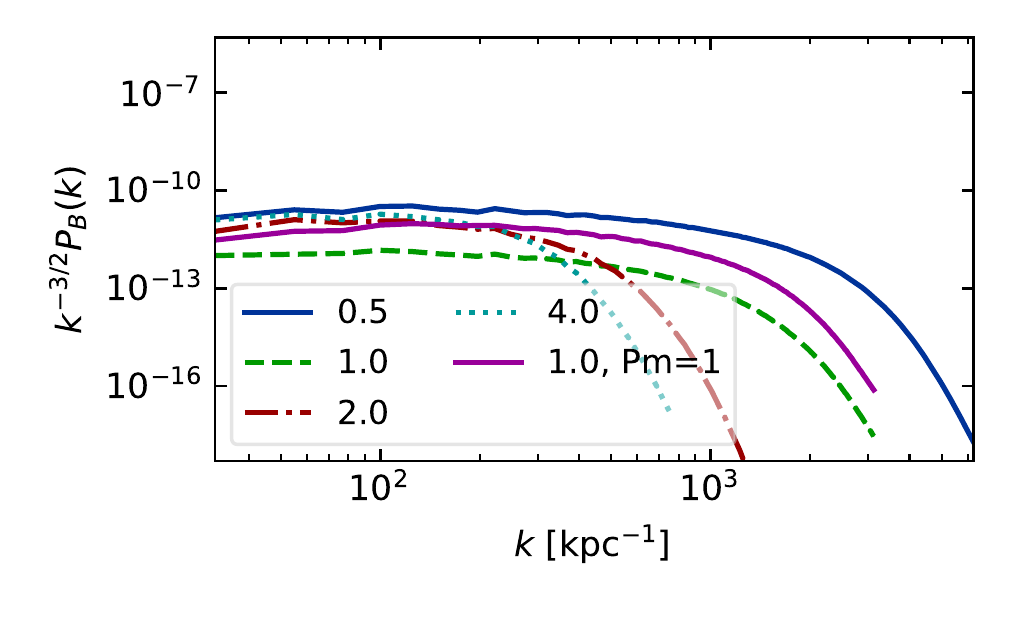}
  \includegraphics[trim=0 1.2cm 0 0.2cm,clip=True,width=0.48\textwidth]{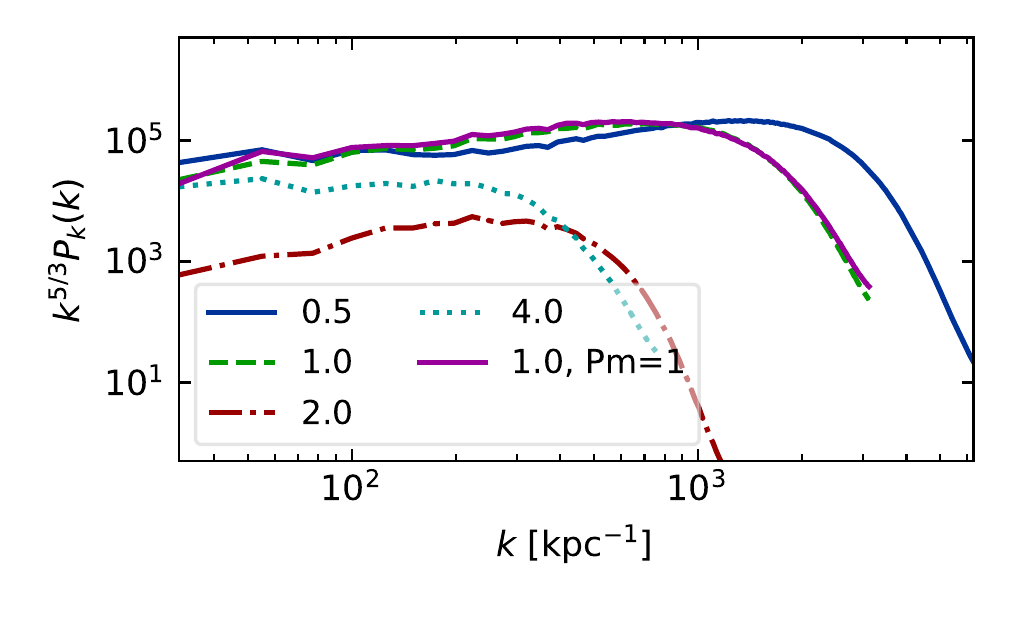}\\
  \includegraphics[trim=0 0.5cm 0 0.2cm,clip=True,width=0.48\textwidth]{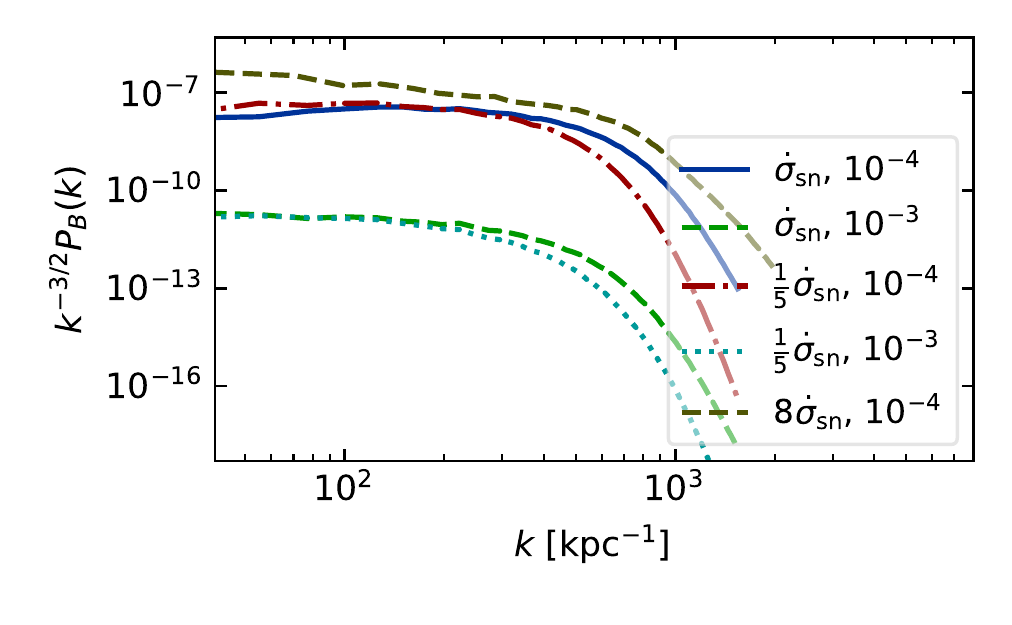}
  \includegraphics[trim=0 0.5cm 0 0.2cm,clip=True,width=0.48\textwidth]{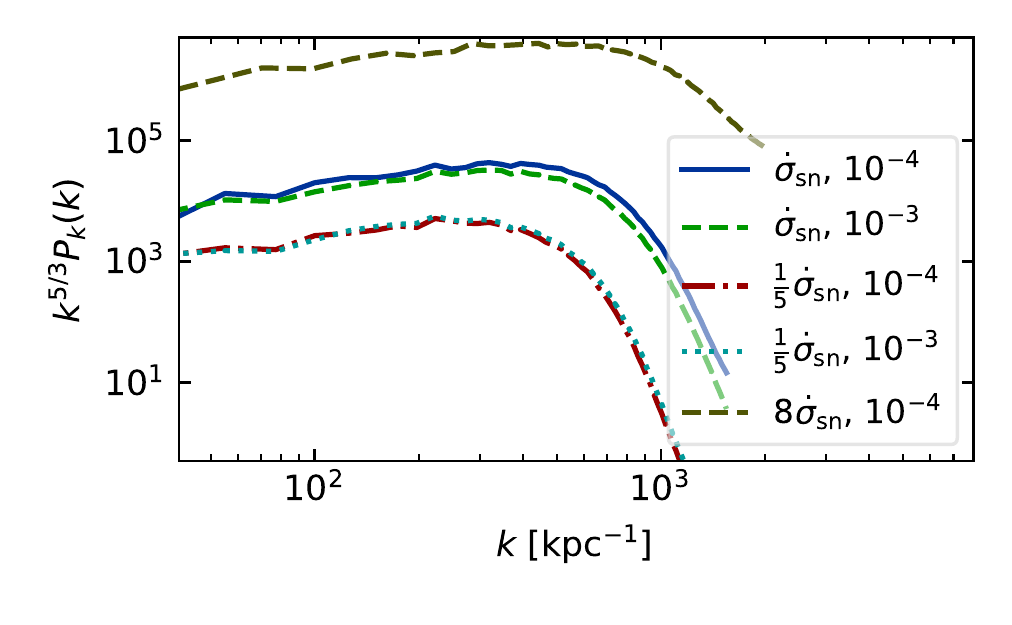}
  \begin{picture}(0,0)(0,0)
    \put( -65,478){{\sf{$\eta=0$}}}
    \put(-315,478){{\sf{$\eta=0$}}}
    \put( -65,466){{\sf{$\dot\sigma=0.2\SNr$}}}
    \put(-315,466){{\sf{$\dot\sigma=0.2\SNr$}}}
    \put( -65,358){{\sf{$\eta=10^{-4}$}}}
    \put(-315,358){{\sf{$\eta=10^{-4}$}}}
    \put( -65,346){{\sf{$\dot\sigma=0.2\SNr$}}}
    \put(-315,346){{\sf{$\dot\sigma=0.2\SNr$}}}
    \put( -65,240){{\sf{$\eta=10^{-3}$}}}
    \put(-315,240){{\sf{$\eta=10^{-3}$}}}
    \put( -65,228){{\sf{$\dot\sigma=0.2\SNr$}}}
    \put(-315,228){{\sf{$\dot\sigma=0.2\SNr$}}}
    \put(-325,119){{\sf{$\dx=1.56,2$}}}
    \put( -75,119){{\sf{$\dx=1.56,2$}}}
    \put(-492,482){{\sf\bf{(a)}}}
    \put(-492,358){{\sf\bf{(b)}}}
    \put(-492,240){{\sf\bf{(c)}}}
    \put(-492,125){{\sf\bf{(d)}}}
  \end{picture}
\caption{
 Compensated {magnetic (left) and kinetic (right)} energy spectra.
 Compensation is {for} the Kazantsev spectrum $k^{3/2}$ (left) or the
 Kolmogorov spectrum $k^{-5/3}$ (right).
 (a)--(c): {Samples at $t=27.5\Myr$ for $\dx$ {given
    in the legend} for
 resistivity $\eta$ and supernova rate $\dot{\sigma}$ as indicated
 {on each panel.  {Field strengths reached differ between models} at
 this time (see Fig.~\ref{fig:eb-res}). Viscosity}
 $\nu=0$, except $\nu=10^{-3}$ for models identified by Pm.
 (d): Samples taken {at different times but} similar field strength at
 $\eta$ {and} $\dot\sigma$ as indicated {in the legend}, {with the
 highest supernova rate being the $\dx = 1.56$ run from the comparison with
 \BKM\ }and $\dx=2$ otherwise.}
 \label{fig:3power}}
\end{figure*}

\section{Results} \label{sec:results}

\subsection{{Resolution and convergence}} \label{sec:conv}
{Figure~\ref{fig:eb-res} shows
 that numerical diffusion still dominates at studied
 resolutions for resistivity  $\eta = 10^{-4}$, as can be seen from the
 increasing speed of the SSD with resolution, but that a converged SSD solution
 emerges at  $\eta = 10^{-3}$ for parsec resolution.}
 Saturation at around 5\% of $\overline{e_K}$ appears to be a well-converged
 result.
 {The $\eta = 10^{-3}$ models show} false convergence
 \citep{FMA91} of solutions with similar magnetic energy decay at $\dx\geq2$.
 {We note that strong fluctuations in the characteristics of the flow
 {occur at the} low $\dot\sigma$ {that we choose to avoid thermal runaway
 \citep{LOCBN15},} with thermal phases occupying changing fractional volumes
 \citep[e.g.][]{gatto2015} and hosting SSD instabilities with different 
 thresholds and growth rates.
 }

 {We ran models to reproduce the results of \BKM, which adopt their choice
 $\dot\sigma = 8\SNr$. In our fiducial runs we use a lower value of $0.2\SNr$} 
 to preserve multiphase thermal structure.
 The higher $\dot\sigma$ rapidly drives thermal runaway resulting in high
 temperatures $T>10^7$\,K.
 The growth rate is faster {than in our fiducial models} (Figure\,\ref{fig:eb-res}\,(c)), reflective of the
 single phase kinematics and more persistent forcing rate, yet {still}
 saturating at about 5\% of $\overline{e_K}$.
 {At equivalent resolution, the sixth-order Pencil Code has far lower diffusion
 than the second-order Godunov code used by \BKM. As a result, we find faster
 growth at equivalent resolution.}
{
         Figure~\ref{fig:eb-res}(d) shows that kinetic energy fluctuates
         around a stationary mean in our models, with higher SN rate $\dot{\sigma}$
         producing higher kinetic energy,
 less intermittency in the energy, and less erratic growth in the
 dynamo.} 
 
 We can also examine the
      {kinetic and magnetic}
   energy spectra (Figure~\ref{fig:3power}).
   The kinetic spectra
   for $\dx\leq1$ agree well at 
 all scales above the viscous cutoff, which appears, as expected, at lower $k$ for
 $\dx=1$ than $\dx=0.5$.
 By contrast the kinetic spectra for $\dx\geq2$ differ and exhibit significant
 energy losses at all scales,    
 indicating only solutions for $\dx\lesssim1$ have converged.

\begin{figure*}
  \includegraphics[trim=0.5cm 0.0cm 0.3cm 0.0cm, clip=true,width=\columnwidth]{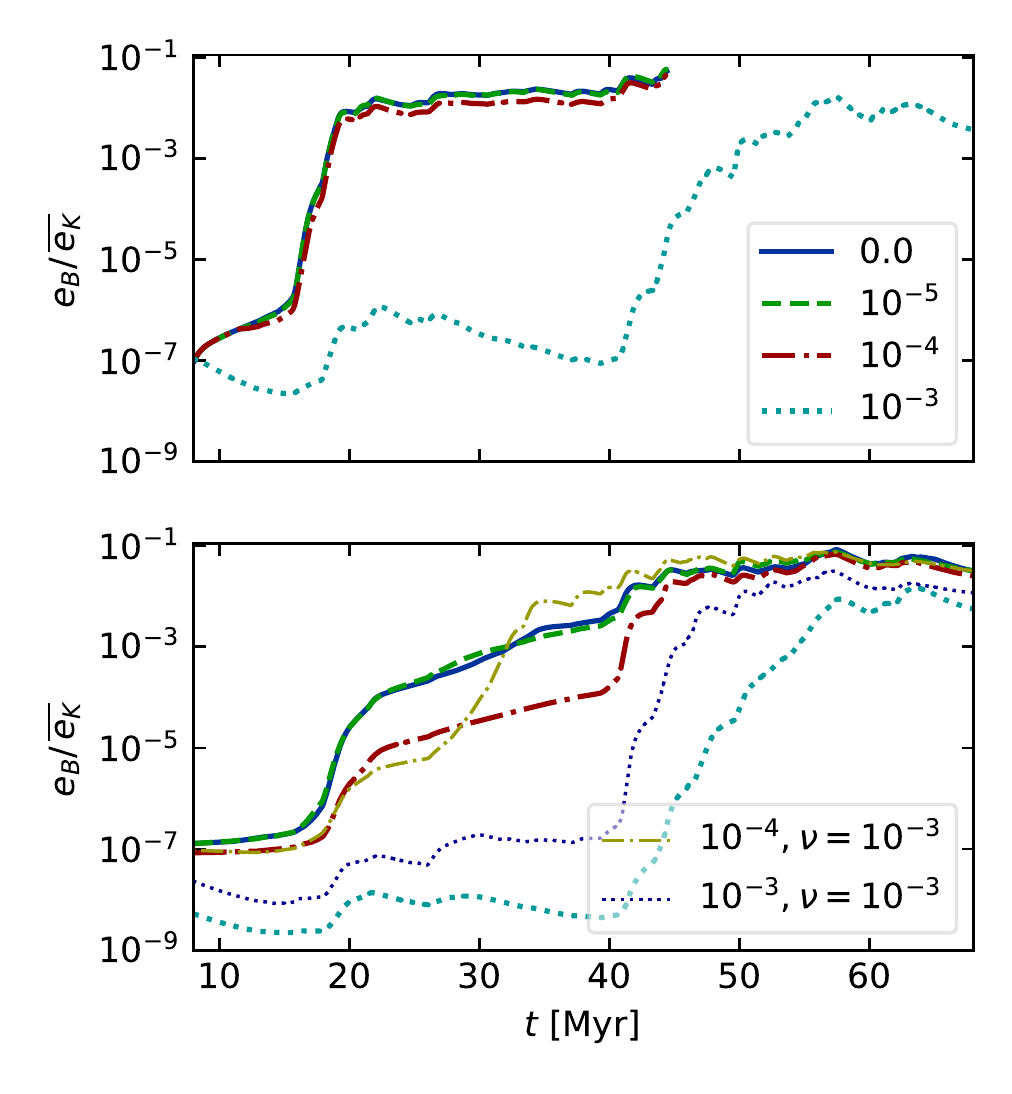}
  \includegraphics[trim=0.5cm 0.0cm 0.3cm 0.0cm, clip=true,width=\columnwidth]{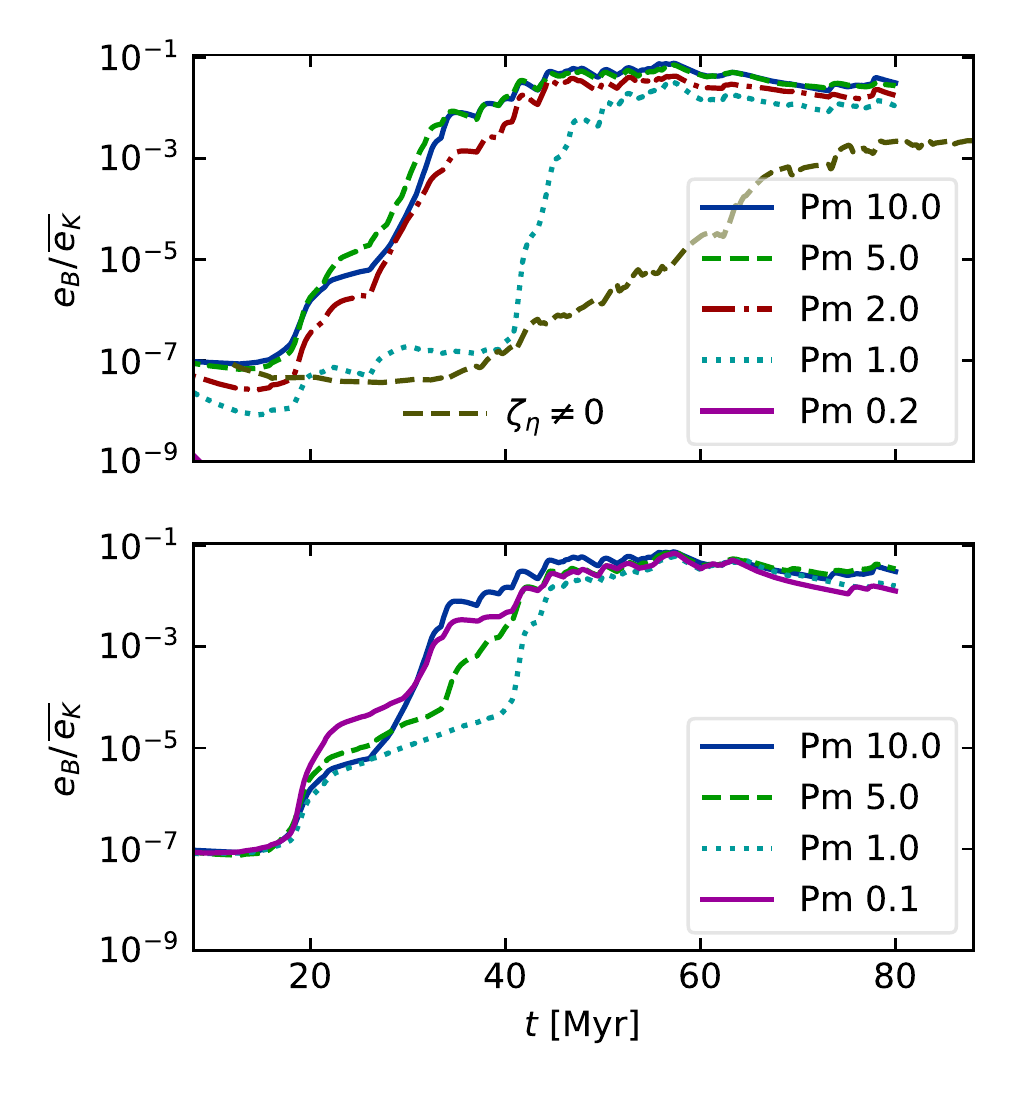}\\
  \includegraphics[trim=0.5cm 0.0cm 0.3cm 0.0cm, clip=true,width=\columnwidth]{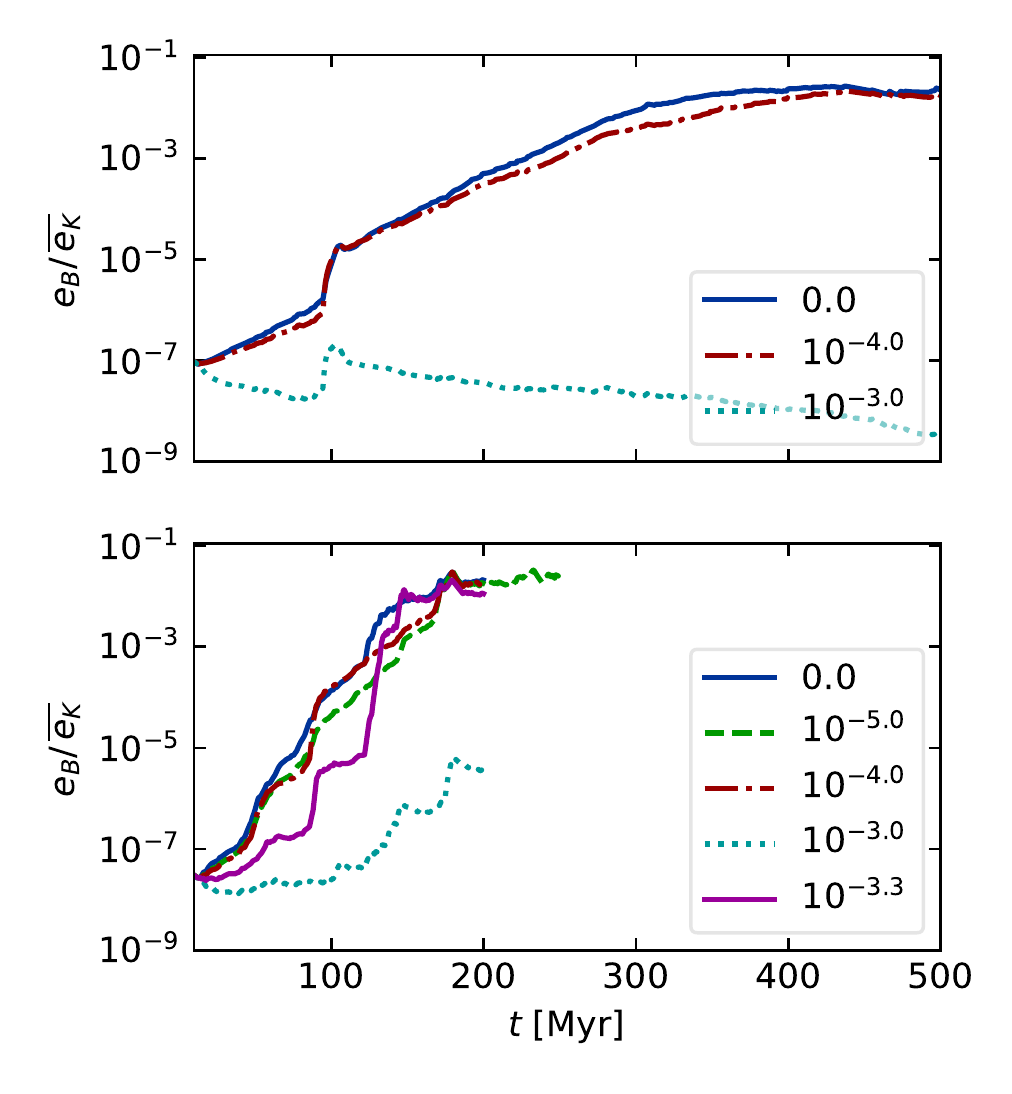}
  \includegraphics[trim=0.5cm 0.0cm 0.3cm 0.0cm, clip=true,width=\columnwidth]{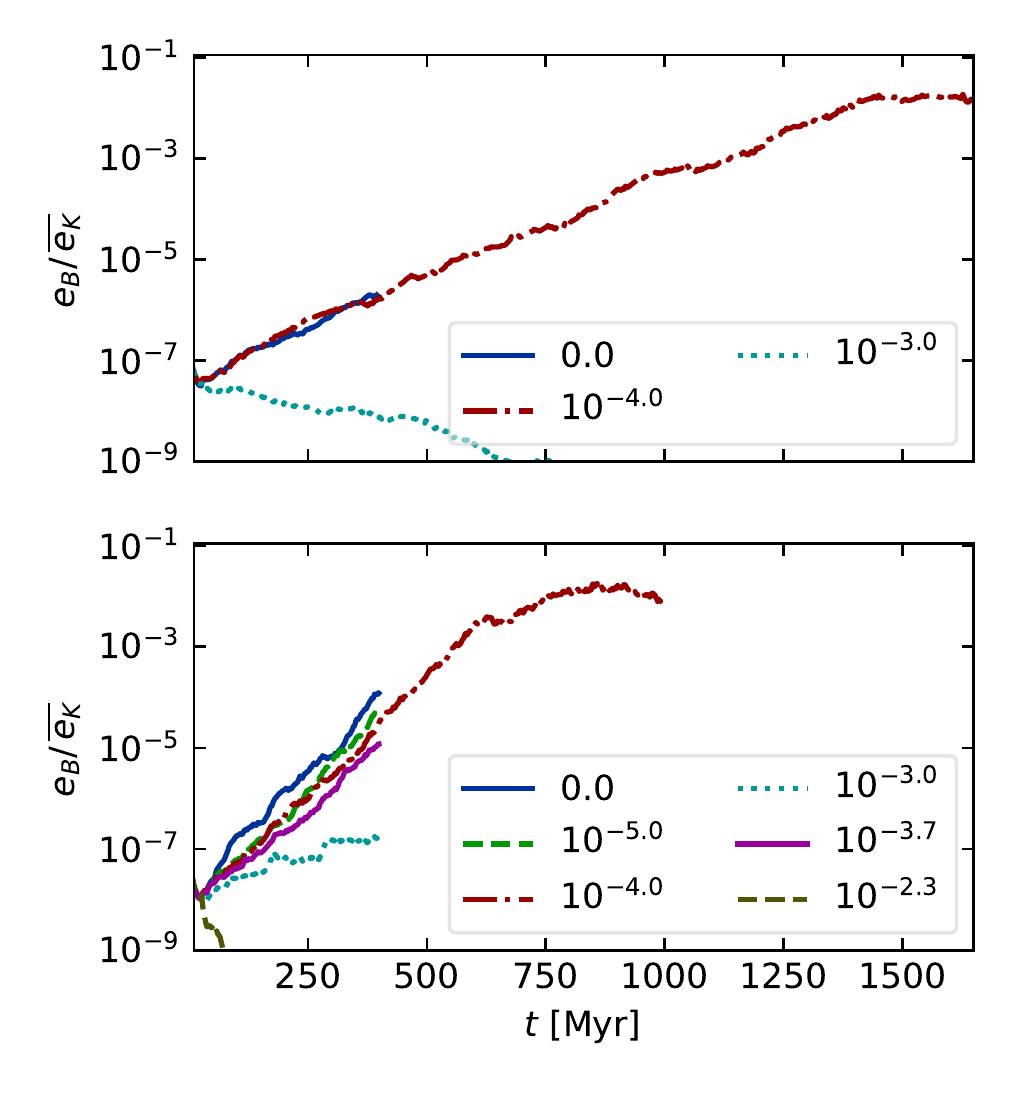}
  \begin{picture}(0,0)(0,0)
    \put(-452,544){{\sf{$\delta x=0.5$}}}
    \put(-452,531){{\sf{$\dot\sigma=\frac{1}{5}\SNr$}}}
    \put(-205,544){{\sf{$\delta x=1$}}}
    \put(-205,531){{\sf{$\nu=10^{-3}$}}}
    \put(-205,518){{\sf{$\dot\sigma=\frac{1}{5}\SNr$}}}
    \put(-452,415){{\sf{$\delta x=1$}}}
    \put(-452,402){{\sf{$\dot\sigma=\frac{1}{5}\SNr$}}}
    \put(-205,415){{\sf{$\delta x=1$}}}
    \put(-205,402){{\sf{$\eta=10^{-4}$}}}
    \put(-205,389){{\sf{$\dot\sigma=\frac{1}{5}\SNr$}}}
    \put(-452,260){{\sf{$\delta x=2$}}}
    \put(-452,247){{\sf{$\dot\sigma=\frac{1}{5}\SNr$}}}
    \put(-205,260){{\sf{$\delta x=4$}}}
    \put(-205,247){{\sf{$\dot\sigma=\frac{1}{5}\SNr$}}}
    \put(-452,133){{\sf{$\delta x=2$}}}
    \put(-452,120){{\sf{$\dot\sigma=\SNr$}}}
    \put(-205,133){{\sf{$\delta x=4$}}}
    \put(-205,120){{\sf{$\dot\sigma=\SNr$}}}
    \put(-450,455){{\sf\bf{(a1)}}}
    \put(-450,330){{\sf\bf{(a2)}}}
    \put(-450,170){{\sf\bf{(c1)}}}
    \put(-450, 45){{\sf\bf{(c2)}}}
    \put(-205,455){{\sf\bf{(b1)}}}
    \put(-205,330){{\sf\bf{(b2)}}}
    \put(-205,170){{\sf\bf{(d1)}}}
    \put(-205, 45){{\sf\bf{(d2)}}}
  \end{picture}
\caption{
Magnetic energy density $e_B$ normalized by the time-averaged kinetic
energy $\overline{e_K}$ for values given in each panel of resolution $\dx$ and
SN rate $\dot\sigma$.
Time axes {extend for $\dx\geq2$} to accommodate SSD saturation.
{Captions indicate resistivity $\eta$, with 
{viscosity}
$\nu=0$, unless indicated otherwise or where $\Pm=\nu/\eta$ is varied with
$\nu$ fixed (b1) or $\eta$ fixed (b2). {In (b1) the run with $\zeta_\eta=
0.55$ has $\Pm=10$. All other models have $\zeta_\eta=0$.}
Line styles in (a1) also apply in (a2).}
\label{fig:eb-nu}}
\end{figure*}
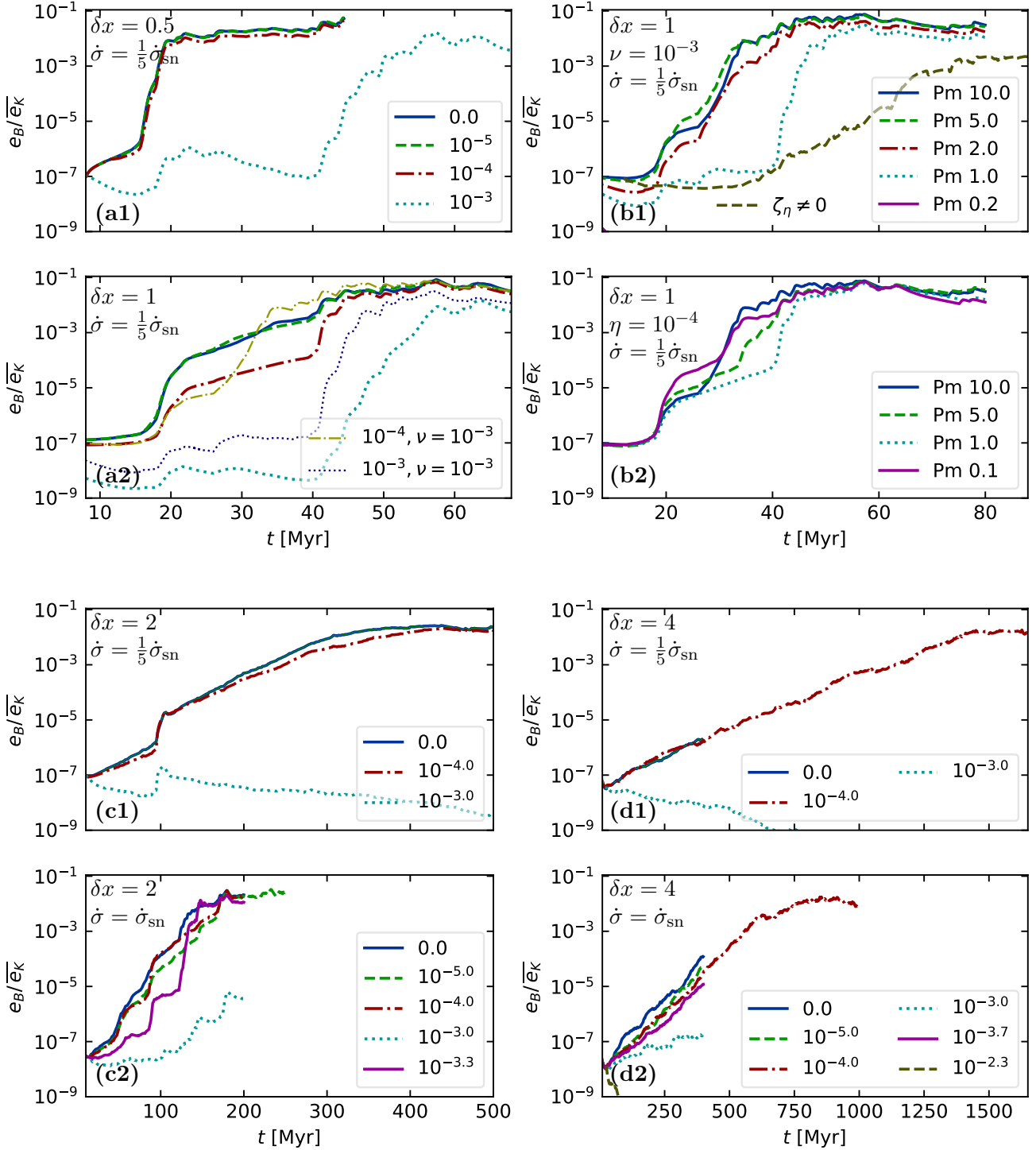

 {The addition of viscosity $\nu=10^{-3}$, indicated by the magenta curves in
 panels (b) and (c), {makes little difference to the shape of the
   magnetic or kinetic energy spectra.  The magnitude of the magnetic
   energy spectrum increases somewhat with the addition of viscosity,
   as can also be seen from comparing dotted light and dark blue
   lines in Figure\,\ref{fig:eb-nu}\,(a2).}
   }

 {We have shown that SSD turbulence {converges} for $\dx\leq1$.
 Underresolving SN driven turbulence results in a significant loss of energy
 at all scales. {The}
 SSD for $0.2\SNr\leq \dot\sigma \leq 8\SNr$ saturates in the ISM at about
 $5\%\overline{e_K}$ and grows more rapidly with increasing SN rate.
 } 

\subsection{{Effective resistivity and Prandtl number}} \label{sec:eta}

 {To understand the role of {physical} {resistivity} $\eta$
   and {viscosity} $\nu$ on the
   SSD,
 we need to determine the value at each resolution where they
   exceed numerical diffusion in strength.}
 Figure~\ref{fig:eb-nu} shows that
 {a physical resistivity of $\eta=10^{-5}$ {(panels a1, a2)}
 makes no impact on field
   growth at $\dot\sigma = 0.2 \SNr$, while $\eta=10^{-3}$ clearly
   dominates over numerical resistivity at all resolutions.  The exact
   value of the minimum physical resistivity does seem to vary not
   just with $\dx$ but also with $\sigma$, as can be seen by
   comparison of the $\eta=10^{-4}$ and {$10^{-3}$ cases (panels c1 to d2)}.
 }

 {When we consider $e_B$ for the models {with} only numerical
 viscosity
 (Fig\,\ref{fig:eb-nu} a, c, d), $\eta\geq10^{-3}$ initially appears
 sufficient to suppress SSD.
 At low resolution this remains so for $\dot\sigma=0.2\SNr$ (panels c1 and d1),
 apart from a transitory surge near 100\,Myr for $\dx=2$.
 However, for $\dot\sigma=\SNr$ within 100\,Myr SSD is evident.
 Only, $\eta=0.005$ dampens SSD (panel d2).}

 {The kinetic energy spectra in Figure\,\ref{fig:3power}} may
 show the resolution of this contradiction.
 They display a bottleneck effect \citep{Falkovich94,HBD03}, an energy cascade
 less efficient than $k^{-5/3}$
 leading to an accumulation of power and then rapid dissipation at high $k$.
 This bottleneck shifts to lower $k$ as $\dx$ {increases}
 {(panels a--c) {or $\dot\sigma$ decreases} (panel d).}
 {The deeper into the magnetic energy spectrum this peak extends, the more 
 scales available for transfer to magnetic energy and the more
 efficient the SSD.
 The critical resistivity above which SSD is suppressed, therefore, 
 increases with $\dot\sigma$, within the range considered. 
 Even at $\dot\sigma=0.2$, for $\eta=10^{-3}$ and $\dx\leq1$ SSD occurs after
 20 -- 40\,Myr.}

\begin{figure}
  \includegraphics[trim=0.5cm 0.2cm 0.3cm 0.0cm, clip=true,width=\columnwidth]{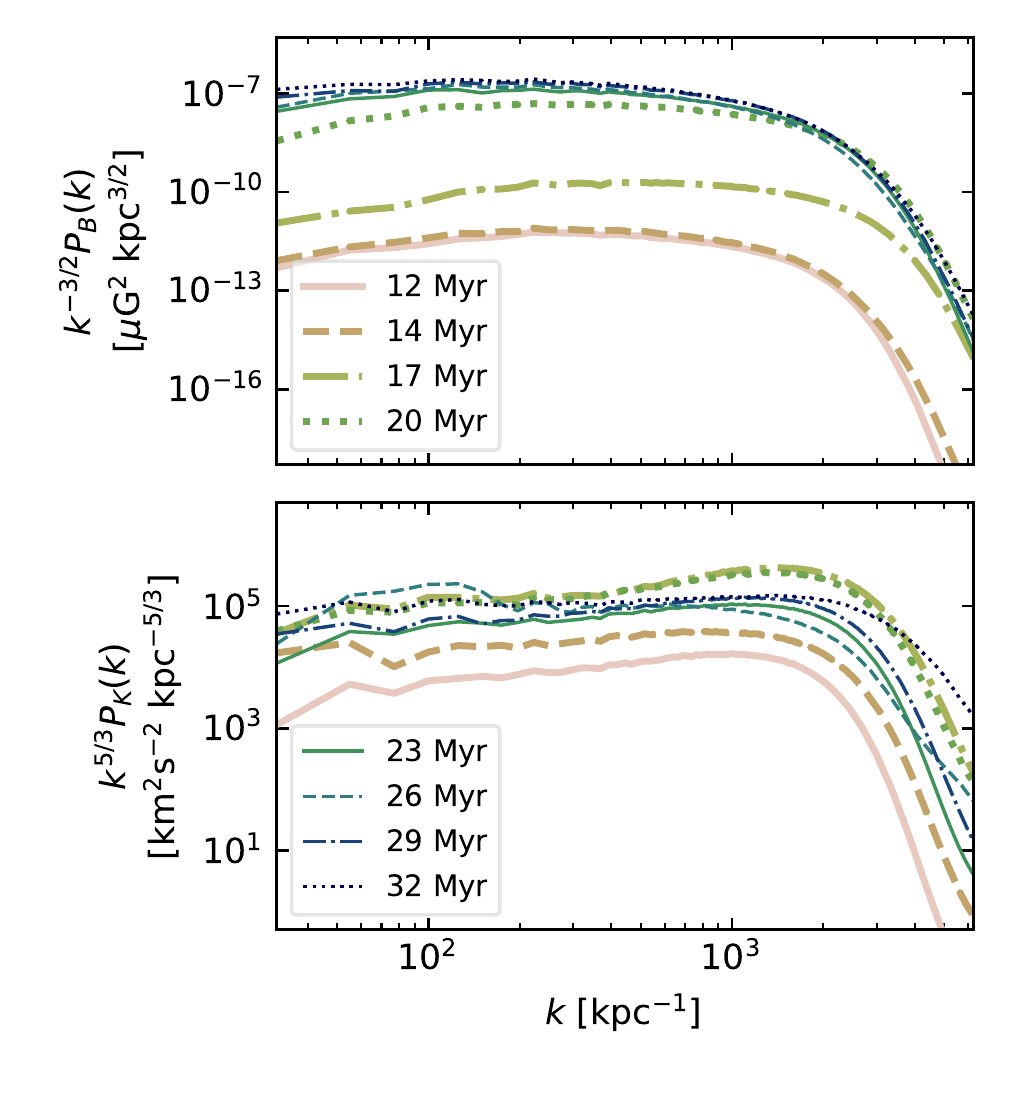}
  \includegraphics[trim=0.5cm 0.5cm 0.3cm 0.0cm, clip=true,width=\columnwidth]{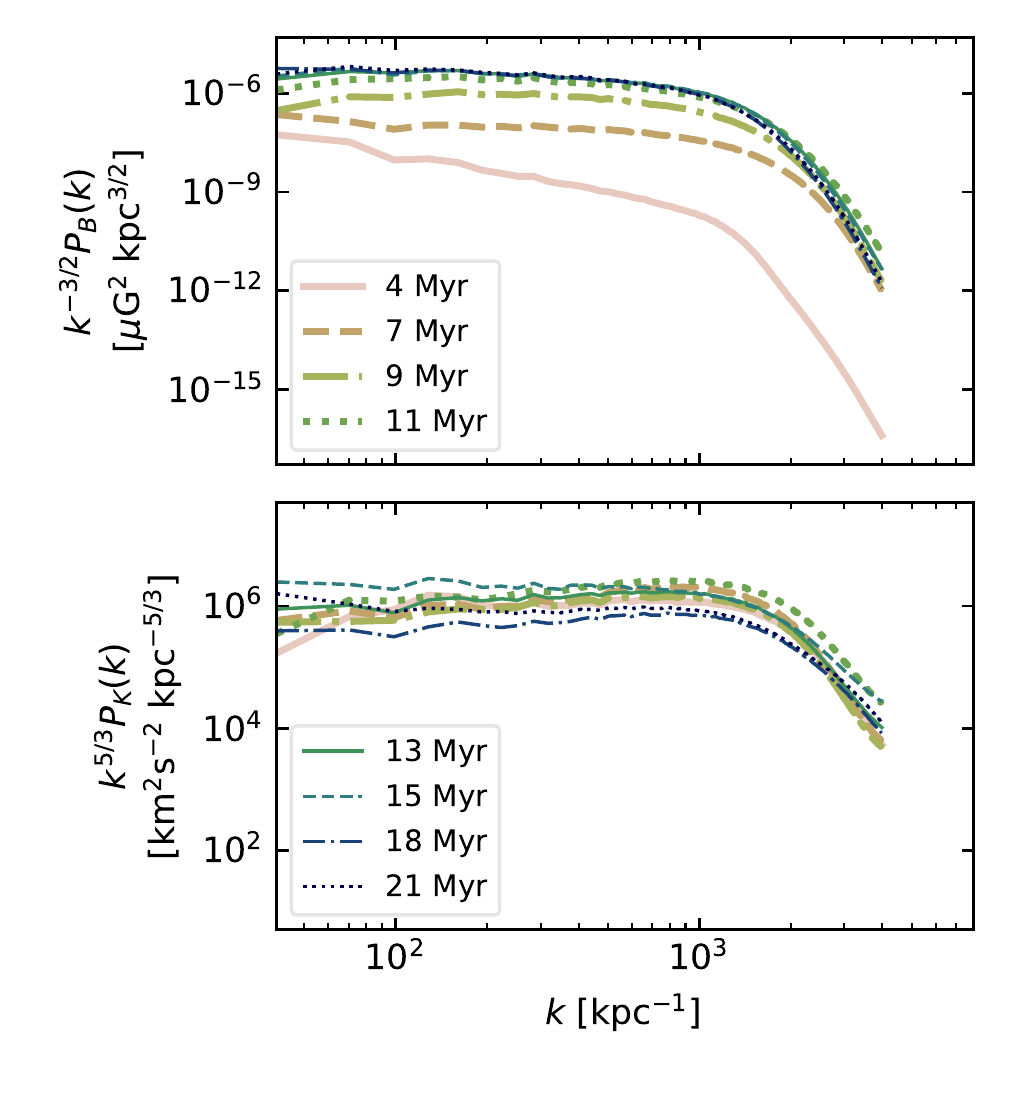}
  \begin{picture}(0,0)(0,0)
    \put(125,483){\sf{{$\delta x=0.5 $}}}
    \put(125,468){\sf{{$\dot\sigma=\frac{1}{5}\SNr$}}}
    \put(125,455){\sf{{$\nu=0$}}}
    \put(125,366){\sf{{$\delta x=0.5 $}}}
    \put(125,353){\sf{{$\dot\sigma=\frac{1}{5}\SNr$}}}
    \put(125,340){\sf{{$\nu=0$}}}
    \put(125,175){\sf{{$\nu=10^{-3.6}$}}}
    \put(125, 50){\sf{{$\nu=10^{-3.6}$}}}
    \put(125,201){\sf{{$\delta x=0.78$}}}
    \put(125, 76){\sf{{$\delta x=0.78$}}}
    \put(125,188){\sf{{$\dot\sigma=          8\SNr$}}}
    \put(125, 63){\sf{{$\dot\sigma=          8\SNr$}}}
    \put(210,540){{\sf\bf{(a1)}}}
    \put(210,415){{\sf\bf{(a2)}}}
    \put(210,260){{\sf\bf{(b1)}}}
    \put(210,135){{\sf\bf{(b2)}}}
  \end{picture}
\caption{
Compensated spectra as in Figure\,\ref{fig:3power} of magnetic (a1, b1)
and kinetic (a2, b2) energy at times given in the legends
  (combined for each pair) for $\dx${, $\dot\sigma$ and $\eta$ indicated.
Resistivity is $\eta=10^{-4}$ with (b1,b2) a comparison to $\BKM$.}
 \label{fig:4power}}
\end{figure}

 {Resistivity contributes to Rm, which is expected to control the onset
 of the SSD and affect growth rate.
 We therefore anticipate that lower $\eta$ would correlate with higher growth
 rate \citep{Sch07}.
 This mainly is the case when we compare models with $\dx\leq1$ at 
 concurrent stages in their evolution.
 However, in Figure\,\ref{fig:eb-nu}\,(c2) there are some anomolous patterns, 
 where higher $\eta$ models overtake lower $\eta$ models, e.g., at 80\,Myr.
 To explore this further we include experiments with $\dx=1$ and $\nu>0$, and
 examine the effect of Pm on the SSD (Figure\,\ref{fig:eb-nu}\,b1, b2).
 We identify each model by $\Pm=\nu/\eta$, but due to the inclusion of 
 shock and hyper diffusivities, the effective Pm and, indeed, Rm vary 
 substantially across space and time.}
 {In b1 we include one run with shock resistivity, $\zeta_\eta\neq0$
 (olive, dashed), which is referenced in Figure\,\ref{fig:eb-slice}.
 The dynamo is slower and saturates lower than the comparative model $\Pm=10$
 (blue, solid).
 This is consistent with more efficient dissipation of compressed field.
}

 {Plotted in panel b2, where we fix $\eta=10^{-4}$ and vary $\nu$,
 initial growth of $e_B$ is faster for 
 $\Pm=0.1$ than for higher values.}
 This is a regime {\em less} conducive to exciting the SSD than the high $\Pm$
 regime typical of the ISM \citep{HBD04}.
 {A plausible explanation may be that the higher fluid Reynolds number, Re,
 could facilitate the dynamo.
 We therefore set a physical viscosity $\nu=10^{-3}$ and vary $\eta$. 
 Plotted in panel (b1) the growth rates mainly
 conform to our expectations, except for $\Pm=5$ between 20 and 40\,Myr.
 We confirm that $\eta\geq10^{-2.3}$ suppresses SSD at $\dot\sigma=0.2\SNr$.
 While the saturation level is insensitive to Pm, with $\nu$ fixed (panel
 b2), the saturation level increases with Pm for $\eta$ fixed (panel b1),
 indicating saturation level is sensitive to Re.
 We also include two of these plots in panel (a2) for comparison to $\nu=0$.
 Comparing the kinetic energy spectra (magenta) models in 
 Figure\,\ref{fig:3power}\,(b2, c2), $\nu$ alters very little.}
 
 {We have shown that the critical resistivity for SSD in the ISM with a
 low SN rate is $10^{-2.3}>\eta_{\rm crit}>10^{-3}$ and
 that this increases with increasing
 $\dot\sigma$ within the range considered.
 Although higher Rm and Pm generally increase growth rate
 and saturation
 in line with theoretical expectations,
 there is considerable variation, likely due to intermittency in
   the multiphase ISM.}

\subsection{{Tangling of the imposed field}} \label{sec:Balsara}

{We now examine whether the field growth seen in our models could
  be due to tangling.
  In Sect.~\ref{sec:ssd-tang} we argued that tangling should produce linear
  growth rate, with dissipation dominating scales below the forcing range.
  Conversely, an SSD leads to exponential growth and a Kazantsev cascade
  extending below the forcing scale.}

 SN-driven turbulence does not have a {single} forcing scale, 
 because of {explosions randomly located in} the heterogeneous ISM.
 {Instead, the} forcing {is} distributed at scales {of roughly 
 60--200~pc
   \citep{joung2006,avillez2007,HSSFG17}, or $k \sim 30$--105~kpc$^{-1}$}.

 {In {Figures~\ref{fig:eb-res} and~\ref{fig:eb-nu}} we indeed
 demonstrate strong exponential growth over multiple orders of magnitude
 {for sufficiently low resistivity, at varying supernova} rate $\dot\sigma$,
 numerical resolution $\dx$, and physical viscosity $\nu$. 
 {Apparently l}inear growth {occurs only} with high physical resistivity.}

 {We now turn to the power spectra. Figure\,\ref{fig:4power} shows
 compensated spectra over time during {intervals} that}
 {span epochs with distinct rates of SSD  growth followed by saturation.}
 The compensated magnetic energy spectra in Figure\,\ref{fig:4power}\,(a1, b1)
 have {peaks conforming to the end of the Kazantsev range.}

 {For ${\dot\sigma=0.2\SNr}$ up to 14\,Myr this peak is at
 $k{\simeq200}\kpc^{-1}$ while the
 SSD grows {slowly}.
 During accelerated growth the Kazentsev range extends to}
 $k\gtrsim{700}\kpc^{-1}$, above the forcing scale and consistent
 with SSD {as shown in the uniform, isothermal model}
 (Fig.\,\ref{fig:tangling}b).
 The peak contracts upon saturation to $k<{200}\kpc^{-1}$, consistent with no
 {further} dynamo (Fig.\,\ref{fig:tangling}c).

 In Figure\,\ref{fig:4power}\,(b1) {until 7\,Myr there is no Kazantsev range
 and the peak energy increases as $k\rightarrow0$, a signature of tangling of
 the imposed field.
 However, as the magnetic field grows much larger than the imposed field,
 this signature disappears and the peak shifts to high $k${, suggesting a healthy SSD}.}
 
 {We have demonstrated that the magnetic field amplification in \BKM\ is
 due to SSD. 
 Tangling of the imposed field is {initially} present, but is
 dominated by SSD. {O}ur other models {with only a weak seed field} 
  confirm {that} an imposed field is not required.
 }

\section{Conclusions}\label{sec:conc}

 {Through the most extensive resolution and parameter study to date, we
   demonstrate in this Letter that SSD
{likely occurs easily}
 in the ISM.
 The critical resistivity is $0.005>\eta_{\rm crit}\gtrsim0.001\kpc\kms$ for 
 supernova rate $\dot \sigma=0.2\SNr$ and increasing over the 
 range considered $\dot \sigma\in(0.2\SNr,8\SNr)$.
 The SSD
 saturates at about 5\% of the equipartition kinetic energy
 density.
 This level is insensitive to Pm, but increases with increasing Re.}
 We find that the conventional approach from dynamo theory of categorising the 
 turbulence according to Rm based on a forcing scale $\ell$, mean random
 velocity $u_{\rm rms}$ and resistivity $\eta$ is inadequate for such a
 complicated system.

 We show that simulations with insufficient resolution can appear to
 converge to a false solution lacking dynamo activity
 (Fig.\,\ref{fig:eb-res}b). This can occur because these simulations are not
 scale independent. 
 The SN energy input and the physically motivated ISM cooling processes impose
 length and time scales that must be adequately resolved.
 {We obtain convergent results for SSD with grid resolution
 $\dx\lesssim1$.}

{We confirm,
  by comparing models with and without an imposed magnetic field,
 that the field amplification obtained in SN-driven ISM turbulence by
 \citet{BKMM04}
 was evidence of an SSD and not only due to tangling of their imposed field.}
 A seed field of less than 1~nG can be amplified to saturation at microgauss
 levels within about 10\,Myr (Figure\,\ref{fig:eb-res}). 

 \citet{Gressel:2008} and \citet{GE20} have $\dx=8.3$ and $6.7\pc$,
 respectively, and $\eta\simeq10^{-2.2}\kpc\kms$, which appears to exclude
 an SSD.
 \citet{Gent:2013a} with $\dx=4\pc$ applied $\eta\simeq10^{-3.1}\kpc\kms$,
 which would support SSD for $\dot\sigma\simeq\SNr$.
 {We can now construct LSD experiments to explore how SSD impacts the 
onset of LSD, critical $\Omega$, and dependence on $\dot\sigma$.}   

\acknowledgments
 {We thank O. Gressel and D. Elstner for discussions inspiring this work,
 and the anonymous referee for comments producing substantial improvement of
 the presentation.}
 FAG and MJK acknowledge support from the Academy of Finland
 ReSoLVE Centre of Excellence (grant 307411) and the ERC
 under the EU's Horizon 2020 research and innovation
 programme (Project UniSDyn, grant 818665) and computational
 resources from CSC–IT Center for Science, Finland, under Grand
 Challenge GDYNS Project 2001062. 
 M-MML was partly supported by US NSF grant AST18-15461.

\software{Pencil Code \citep{brandenburg2002,Pencil-JOSS}}

\bibliographystyle{aasjournal}

\end{document}